\documentclass[12pt]{article}

\makeatletter
 \@addtoreset{equation}{section}
 
\makeatother
\usepackage{amssymb}
\usepackage{amsbsy}
\usepackage[dvips]{graphicx}

\begin{document}
%%%%%%%%%%% Title Page %%%%%%%%%%%%%%%%%%%%

\begin{titlepage}

\renewcommand{\thefootnote}{\fnsymbol{footnote}}

\begin{flushright}
\begin{tabular}{l}
UTHEP-599\\
RIKEN-TH-176\\
\end{tabular}
\end{flushright}

\bigskip

\begin{center}
{\Large \bf 
Light-cone Gauge NSR Strings\\
in Noncritical Dimensions
\\}
\end{center}

\bigskip

\begin{center}
%% AUTHORS
{\large Yutaka Baba}${}^{a}$\footnote{e-mail:
        ybaba@riken.jp},
{\large Nobuyuki Ishibashi}${}^{b}$\footnote{e-mail:
        ishibash@het.ph.tsukuba.ac.jp},
{\large Koichi Murakami}${}^{a}$\footnote{e-mail:
        murakami@riken.jp}
\end{center}

\begin{center}
${}^{a}${\it 
Theoretical Physics Laboratory, RIKEN,\\
Wako, Saitama 351-0198, Japan}
\end{center}
\begin{center}
${}^{b}${\it
Institute of Physics, University of Tsukuba,\\
Tsukuba, Ibaraki 305-8571, Japan}\\
\end{center}

\bigskip

\bigskip

\bigskip

\begin{abstract}
Light-cone gauge NSR string theory in noncritical dimensions should
correspond to a string theory with a nonstandard longitudinal
part. 
Supersymmetrizing the bosonic case \cite{Baba:2009ns}, 
we formulate a superconformal worldsheet theory 
for the longitudinal variables $X^{\pm},\psi^{\pm}$. 
We show that
with the transverse variables and the ghosts combined, 
it is possible to construct a nilpotent BRST charge. 
\end{abstract}

\setcounter{footnote}{0}
\renewcommand{\thefootnote}{\arabic{footnote}}

\end{titlepage}

%%%%%%%%%%%%%%%%%%%%%%%%%%%%%%%%%%%%%%%%%%%%%
\section{Introduction}

In the light-cone gauge formulation of string theory, it is possible
to consider the theory in noncritical space-time dimensions. 
Since the Lorentz invariance is broken in such dimensions, 
it should correspond to a string theory in a Lorentz noninvariant 
background. 
In the previous paper \cite{Baba:2009ns}, 
we consider bosonic string theory and identify
the worldsheet CFT for the longitudinal variables $X^{\pm}$
which corresponds to such a background.  

What we would like to do in this paper is to supersymmetrize 
the results of Ref.~\cite{Baba:2009ns}. 
We propose a superconformal field theory for
the longitudinal variables $X^{\pm},\psi^{\pm}$
which should correspond to the longitudinal part of
the light-cone gauge NSR string 
in $d\ \left(d\ne10\right)$ dimensions.
We show that the superconformal field theory has the right properties
so that we can construct a nilpotent BRST charge with the
transverse variables and ghosts combined. 

This paper is organized as follows. 
In section \ref{sec:pmCFT}, we propose a superconformal field 
theory for $X^{\pm},\psi^{\pm}$. 
We present the energy momentum tensor and the action,
and define the correlation functions. 
In section \ref{sec:superGamma}, 
we evaluate the partition function
$e^{-\Gamma_{\mathrm{super}}}$ of light-cone gauge
NSR strings on the tree super light-cone diagram
for $N$ strings,
which is necessary to calculate the correlation functions
for $X^{\pm},\psi^{\pm}$. 
In section \ref{sec:Correlation}, 
we calculate the correlation functions and
check that the superconformal field theory possesses 
the desired properties.
Section \ref{sec:Conclusions} is devoted to conclusions 
and discussions. 
In appendix~\ref{sec:int-points},
we present properties of the interaction points
on the super light-cone diagram, which are useful
for the computation of $\Gamma_{\mathrm{super}}$.
In appendix~\ref{sec:otherGamma},
we show that $\Gamma_{\mathrm{super}}$
obtained in section \ref{sec:superGamma} is consistent
with the results in Refs.~\cite{Berkovits:1985ji,Berkovits:1987gp}.

%%%%%%%%%%%%%%%%%%%%%%%%%%%%%%%%%%%%%%%%%%%
\section{$X^{\pm}$ CFT\label{sec:pmCFT}}

\subsection{Energy momentum tensor}

In the bosonic case \cite{Baba:2009ns}, 
the energy momentum tensor of the worldsheet
CFT for the longitudinal variables $X^{\pm}$ is given as 
\begin{equation}
T_{X^{\pm}}\left(z\right)
 =  \partial X^{+}\partial X^{-}
    -\frac{d-26}{12}\left\{ X^{+},z\right\} \;,
\label{eq:bosonicT}
\end{equation}
where $\left\{ X^{+},z\right\} $ is the Schwarzian derivative. 

We would like to consider the supersymmetrized version of this theory. 
In order to deal with the operators, it is convenient to introduce 
the supercoordinate 
$\mathbf{z} \equiv \left(z,\theta\right)$ and
the superfields $X^{\pm}\left(\mathbf{z},\bar{\mathbf{z}}\right)$
which can be expanded as
\begin{equation}
X^{\pm}\left(\mathbf{z},\bar{\mathbf{z}}\right)
  =  x^{\pm}+i\theta\psi^{\pm}+i\bar{\theta}\bar{\psi}^{\pm}
     +i\theta\bar{\theta}F^{\pm} \;.
\end{equation}
In this paper we follow the notations of
Refs.~\cite{Friedan1986,Friedan,Aoki:1990yn}
and define
\begin{equation}
D =  \partial_{\theta}+\theta\partial_{z}\;,
\qquad
\mathbf{z}-\mathbf{z}' =  z-z'-\theta\theta'\;.
\end{equation}
The supersymmetric generalization of
eq.(\ref{eq:bosonicT}) will be
\begin{eqnarray}
T_{X^{\pm}}(\mathbf{z}) 
& = & \frac{1}{2}DX^{+}\partial X^{-}+\frac{1}{2}DX^{-}\partial X^{+}
      -\frac{d-10}{4} S(\mathbf{z},\mathbf{X}_{L}^{+}) \;.
\label{eq:TXpm}
\end{eqnarray}
Here 
$\mathbf{X}^{+}_{L}
  =\left( X^{+}_{L} (\mathbf{z}), \Theta^{+}(\mathbf{z}) \right)$;
$X^{+}_{L} (\mathbf{z})$ denotes the holomorphic part
of the superfield $X^{+}(\mathbf{z},\bar{\mathbf{z}})$
and $\Theta^{+}(\mathbf{z})$ is defined as
\begin{equation}
\Theta^{+}\left(\mathbf{z}\right) 
 =  \frac{DX^{+}}{(\partial X^{+})^{\frac{1}{2}}}
        \left(\mathbf{z}\right)\;,
\end{equation}
so that the map 
$\mathbf{z}=\left(z,\theta\right) \mapsto
 \mathbf{X}_{L}^{+}\left(\mathbf{z}\right)
   =\left(X_{L}^{+}\left(\mathbf{z}\right),
          \Theta^{+}\left(\mathbf{z}\right)\right)$
is a superconformal mapping.
$S(\mathbf{z},\mathbf{X}_{L}^{+})$ is 
the super Schwarzian derivative defined as
\begin{eqnarray}
S(\mathbf{z},\mathbf{X}_{L}^{+}) 
& = & \frac{D^{4}\Theta^{+}}{D\Theta^{+}}
       -2\frac{D^{3}\Theta^{+}D^{2}\Theta^{+}}{(D\Theta^{+})^{2}}
\nonumber\\
 & = & -\frac{1}{4}D\Phi\partial\Phi+\frac{1}{2}\partial D\Phi\;,
\end{eqnarray}
where
\begin{equation}
\Phi\left(\mathbf{z},\bar{\mathbf{z}} \right) 
 =  \ln \left( -4
             \left(D\Theta^{+}\right)^{2}\left(\mathbf{z}\right)
             \left(\bar{D}\bar{\Theta}^{+}\right)^{2}
                     \left(\bar{\mathbf{z}}\right)
        \right)\;.
\end{equation}
In the following, 
we will study the superconformal field theory 
with the energy momentum tensor $T_{X^{\pm}}\left(\mathbf{z}\right)$
in eq.(\ref{eq:TXpm}), on the complex plane.  

%%%%%%%%%%%%%%%%%%%%%%%%%%%%%%%%%%%%%%%%%%%%
\subsection{Action and correlation functions}

{}From the energy momentum tensor (\ref{eq:TXpm}), we obtain
the action for the theory:
\begin{eqnarray}
S_{X^{\pm}} 
& = & -\frac{1}{2\pi}\int d^{2}\mathbf{z}
      \left( \bar{D}X^{+} DX^{-} + \bar{D}X^{-} DX^{+} \right)
      +\frac{d-10}{8}\Gamma_{\mathrm{super}}\left[\Phi\right]
\nonumber\\
 & = & -\frac{1}{2\pi}\int d^{2}z
       \left(\partial x^{+}\bar{\partial}x^{-}
             + \partial x^{-}\bar{\partial}x^{+}\right.
\nonumber \\
 &  & \hphantom{ -\frac{1}{2\pi}\int d^{2}z}
      \ \left. +\psi^{+}\bar{\partial}\psi^{-}
         + \psi^{-}\bar{\partial}\psi^{+}
         + \tilde{\psi}^{+}\partial\tilde{\psi}^{-}
         + \tilde{\psi}^{-}\partial\tilde{\psi}^{+}
        \right)
\nonumber \\
 &  & \ +\frac{d-10}{8}\Gamma_{\mathrm{super}}
        \left[\Phi\right]\;.
\end{eqnarray}
Here we have used the measure 
$d^{2}\mathbf{z}
  =d\left(\mathop{\mathrm{Re}} z \right)
   d\left(\mathop{\mathrm{Im}} z\right)
   d\theta d\bar{\theta}$.
$\Gamma_{\mathrm{super}}\left[\Phi\right]$ 
 coincides with the super Liouville action
\begin{equation}
\frac{d-10}{8}\Gamma_{\mathrm{super}}\left[\Phi\right] 
 =  -\frac{d-10}{16\pi}\int d^{2}\mathbf{z}\bar{D}\Phi D\Phi\;,
\label{eq:Liouvilleaction}
\end{equation}
and is responsible for 
the super Schwarzian derivative part of the energy momentum tensor.

In order for the super Liouville action to be well-defined,  
$e^{\Phi}$ should have a nonzero expectation value. 
As in the bosonic case, 
we always consider the theory in
the presence of the vertex operators
$$ e^{-ip_{r}^{+}X^{-}}\left(\mathbf{Z}_{r},
                             \bar{\mathbf{Z}}_{r}\right)$$
for $r=1,\ldots,N$, with $\sum_{r=1}^{N}p_{r}^{+}=0$ 
and $\mathbf{Z}_{r}=\left(Z_{r},\Theta_{r}\right)$.
Thus the quantities in which we are interested are 
the expectation values of functionals 
$F\left[X^{+},X^{-}\right]$ defined as 
\begin{eqnarray}
\lefteqn{ 
  \left\langle F\left[X^{+},X^{-}\right]
    \prod_{r=1}^{N}
       e^{-ip_{r}^{+}X^{-}}\left(\mathbf{Z}_{r},
                                 \bar{\mathbf{Z}}_{r}\right)
           \right\rangle
}
\nonumber \\
&& 
  \equiv \int\left[dX^{+}dX^{-}\right]
   e^{-S_{X^{\pm}}} F\left[X^{+},X^{-}\right]
   \prod_{r=1}^{N}e^{-ip_{r}^{+}X^{-}}
            \left(\mathbf{Z}_{r},\bar{\mathbf{Z}}_{r}\right) \;.
\end{eqnarray}
For the functionals $F\left[X^{+}\right]$ which do not depend 
on $X^{-}$,
it is easy to calculate them up to an overall factor 
as in the bosonic case \cite{Baba:2009ns}
and obtain
\begin{eqnarray}
\lefteqn{
  \left\langle F\left[X^{+}\right]
     \prod_{r=1}^{N}e^{-ip_{r}^{+}X^{-}}
           \left(\mathbf{Z}_{r},\bar{\mathbf{Z}}_{r}\right)
  \right\rangle
}
 \nonumber \\
 &  & \sim F\left[-\frac{i}{2}\left(\rho+\bar{\rho}\right)\right]
      \exp \left(-\frac{d-10}{8}
                 \Gamma_{\mathrm{super}}
                   \left[\ln\left(\left(D\xi\right)^{2}
                            \left(\bar{D}\bar{\xi}\right)^{2}\right)
                   \right]
           \right) \;,
\label{eq:generating}
\end{eqnarray}
where
\begin{equation}
\rho\left(\mathbf{z}\right) 
= \sum_{r=1}^{N}\alpha_{r}\ln\left(\mathbf{z}-\mathbf{Z}_{r}\right)\;,
\qquad 
\xi\left(\mathbf{z}\right) 
 =  \frac{D\rho}{\left(\partial\rho\right)^{\frac{1}{2}}}
    \left(\mathbf{z}\right)\;,
\qquad
\alpha_{r}=2p^{+}_{r}\;.
\label{eq:Mandelstam}
\end{equation}
$\rho\left(\mathbf{z}\right)$ coincides with 
the super Mandelstam mapping 
and $\xi (\mathbf{z})$ is defined so that the mapping 
$\mathbf{z}=\left(z,\theta\right)\mapsto
  \boldsymbol{\rho}=\left(\rho,\xi\right)$
is superconformal. From eq.(\ref{eq:generating}), 
one can see that $X^+(\mathbf{z},\bar{\mathbf{z}})$ has 
the expectation value
$$
-\frac{i}{2}
\left(\rho (\mathbf{z})+\bar{\rho}(\bar{\mathbf{z}})
\right)
\:.
$$

It is convenient to define 
\begin{equation}
\left\langle F \left[X^{+},X^{-}\right] \right\rangle _{\rho} 
 \equiv 
 \frac{\left\langle F \left[X^{+},X^{-}\right]
              \prod_{r=1}^{N}e^{-ip_{r}^{+}X^{-}}
                \left(\mathbf{Z}_{r},\bar{\mathbf{Z}}_{r}\right)
       \right\rangle }
      {\left\langle \prod_{r=1}^{N} e^{-ip_{r}^{+}X^{-}}
            \left(\mathbf{Z}_{r},\bar{\mathbf{Z}}_{r}\right)
       \right\rangle }\;.
\label{eq:corr-rho}
\end{equation}
Various correlation functions can be obtained 
by differentiating eq.(\ref{eq:generating})
with respect to $\alpha_{r}$'s. 
For example, introducing 
\begin{equation}
\rho^{\prime}\left(\mathbf{z}\right)
 \equiv 
 \sum_{r=0}^{N+1} \alpha_{r}
      \ln\left(\mathbf{z}-\mathbf{Z}_{r}\right)\;,
\qquad
\alpha_{N+1}=-\alpha_{0}\;,
\end{equation}
and $\xi' (\mathbf{z})$ accordingly,
we get the one-point function 
$\left\langle 
   DX^{-}\left(\mathbf{Z}_{0}\right)
 \right\rangle _{\rho}$
as
\begin{equation}
\left\langle DX^{-}\left(\mathbf{Z}_{0}\right)
\right\rangle _{\rho} 
 =  \left.  2i\partial_{\alpha_{0}} D_{\mathbf{Z}_{0}}
      \left( -\frac{d-10}{8} 
         \Gamma_{\mathrm{super}}
          \left[ \ln\left( \left(D\xi^{\prime}\right)^{2}
                           \left(\bar{D}\bar{\xi}^{\prime}\right)^{2}
                           \right)
          \right]
      \right) \right|_{\alpha_{0}=0}\;.
\label{eq:onepoint}
\end{equation}
The manipulation is essentially the same as that 
in the bosonic case \cite{Baba:2009ns}
and it is straightforward to obtain
\begin{eqnarray}
& & \left\langle DX^{-}\left(\mathbf{Z}_{0}\right)
         DX^{-}\left(\mathbf{z}\right)\right\rangle _{\rho}
=\left. 2i \partial_{\alpha_{0}} D_{\mathbf{Z}_{0}}
   \left\langle DX^{-}\left(\mathbf{z}\right)
   \right\rangle _{\rho^{\prime}}\right|_{\alpha_{0}=0}
 + \left\langle DX^{-}\left(\mathbf{Z}_{0}\right)
   \right\rangle _{\rho}
   \left\langle DX^{-}\left(\mathbf{z}\right) 
   \right\rangle _{\rho}\ ,
\nonumber \\
& & \left\langle DX^{-}\left(\mathbf{z}\right) F\left[X^{+}\right]
    \right\rangle _{\rho}
 \nonumber \\
 & & \qquad
  = \left\langle DX^{-}\left(\mathbf{z}\right)
    \right\rangle _{\rho}
    F\left[-\frac{i}{2}\left(\rho+\bar{\rho}\right)\right]
   + \int d^{2} \mathbf{z}^{\prime}
     \frac{\theta-\theta^{\prime}}
          {\mathbf{z}-\mathbf{z}^{\prime}}
     \left.\frac{\delta F\left[X^{+}\right]}
                {\delta X^{+}\left(\mathbf{z}^{\prime}\right)}
     \right|_{X^{+}=-\frac{i}{2}\left(\rho+\bar{\rho}\right)}
  \ ,\nonumber \\
& & \left\langle DX^{-}\left(\mathbf{z}\right)
         DX^{-} \left(\mathbf{z}^{\prime}\right)
         F\left[X^{+}\right] \right\rangle _{\rho}
  \nonumber \\
 & & \qquad = 
    \left\langle DX^{-}\left(\mathbf{z}\right)
                 DX^{-}\left(\mathbf{z}^{\prime}\right)
    \right\rangle 
       F\left[-\frac{i}{2}\left(\rho+\bar{\rho}\right)\right]
  \nonumber \\
 & & \qquad \quad \hphantom{=}
   {} +\left\langle DX^{-} \left(\mathbf{z}\right)
        \right\rangle _{\rho}
      \int d^{2}\mathbf{z}^{\prime\prime}
       \frac{\theta^{\prime}-\theta^{\prime\prime}}
            {\mathbf{z}^{\prime}-\mathbf{z}^{\prime\prime}}
       \left.
        \frac{\delta F\left[X^{+}\right]}
             {\delta X^{+}\left(\mathbf{z}^{\prime\prime}\right)}
       \right|_{X^{+}=-\frac{i}{2}\left(\rho+\bar{\rho}\right)}
   \nonumber \\
 & & \qquad \quad \hphantom{=}
    {}+\int d^{2}\mathbf{z}^{\prime\prime}
        \frac{\theta-\theta^{\prime\prime}}
             {\mathbf{z}-\mathbf{z}^{\prime\prime}}
       \left.
         \frac{\delta F\left[X^{+}\right]}
              {\delta X^{+}\left(\mathbf{z}^{\prime\prime}\right)}
       \right|_{X^{+}=-\frac{i}{2}\left(\rho+\bar{\rho}\right)}
       \left\langle DX^{-}\left(\mathbf{z}^{\prime}\right)
       \right\rangle _{\rho}
    \nonumber \\
 & & \qquad \quad \hphantom{=}
    {}+\int d^{2}\mathbf{z}^{\prime\prime}
            d^{2}\mathbf{z}^{\prime\prime\prime}
       \frac{\theta-\theta^{\prime\prime}}
            {\mathbf{z}-\mathbf{z}^{\prime\prime}}
       \frac{\theta^{\prime}-\theta^{\prime\prime\prime}}
            {\mathbf{z}^{\prime}-\mathbf{z}^{\prime\prime\prime}}
       \left.
         \frac{\delta^{2}F\left[X^{+}\right]}
              {\delta X^{+}\left(\mathbf{z}^{\prime\prime}\right)
                \delta X^{+}
                   \left(\mathbf{z}^{\prime\prime\prime}\right)}
        \right|_{X^{+}=-\frac{i}{2}\left(\rho+\bar{\rho}\right)}
      \ , 
\label{eq:corr}
\end{eqnarray}
and so on. In this way we can, in principle, derive 
all the correlation functions from 
the one-point function in eq.(\ref{eq:onepoint}). 
From eq.(\ref{eq:corr}),
one can read off the OPE's
\begin{eqnarray}
X^{+}\left(\mathbf{z},\bar{\mathbf{z}}\right) 
 X^{+}\left(\mathbf{z}^{\prime},\bar{\mathbf{z}}^{\prime} \right)
 & \sim & \mathrm{regular}\; ,
\nonumber \\
DX^{-}\left(\mathbf{z}\right) 
  X^{+}\left(\mathbf{z}^{\prime},\bar{\mathbf{z}}^{\prime} \right)
 & \sim & \frac{\theta-\theta^{\prime}}
               {\mathbf{z}-\mathbf{z}^{\prime}} \; .
\end{eqnarray}
In order to obtain the OPE 
$X^{-}
%\left(\mathbf{z},\bar{\mathbf{z}} \right)
X^{-}
%\left(\mathbf{z}^{\prime}, \bar{\mathbf{z}}^{\prime}\right)
$,
we need the explicit form of 
$\Gamma_{\mathrm{super}}\left[ \ln\left( \left(D\xi\right)^{2}
\left(\bar{D}\bar{\xi}\right)^{2}\right)\right]$, 
which will be denoted as $\Gamma_{\mathrm{super}}$ in the following.

%%%%%%%%%%%%%%%%%%%%%%%%%%%%%%%%
\section{Evaluation of $\Gamma_{\mathrm{super}}$}
 \label{sec:superGamma}

As in the bosonic case, 
even with the expectation value 
$-\frac{i}{2}\left(\rho +\bar{\rho}\right)$ 
for $X^+$, 
the super Liouville action (\ref{eq:Liouvilleaction}) 
is not well-defined
because of the singularities of 
$\left\langle
  \Phi\left(\mathbf{z},\bar{\mathbf{z}}\right)
 \right\rangle_{\rho}$. 
In order to define $\Gamma_{\mathrm{super}}$, 
we should regularize the singularities
and carefully take account of various 
effects~\cite{Mandelstam:1985ww,Berkovits:1985ji,Berkovits:1987gp}.
Here we would like to calculate $\Gamma_{\mathrm{super}}$ 
in the same way as was done in Ref.~\cite{Baba:2009ns}. 
As in the bosonic case,
the form obtained by such a method is more convenient 
for the calculations of the correlation functions.
In appendix \ref{sec:otherGamma}, 
we show that $\Gamma_{\mathrm{super}}$ we get is 
consistent with the results of
Ref.~\cite{Berkovits:1985ji,Berkovits:1987gp}.

\subsection{Procedure}

$\boldsymbol{\rho}$ can be considered as the supercoordinate 
on the super light-cone diagram. 
We would like to obtain $\Gamma_{\mathrm{super}}$
by integrating the variation $\delta\Gamma_{\mathrm{super}}$ 
under the variations of moduli. 
The moduli space of the super light-cone diagram 
corresponds to the space of parameters 
$Z_{r}$ and $\Theta_{r}$ $\left(r=1,\cdots,N\right)$
modded out by the superprojective transformation
\begin{equation}
z  \mapsto 
  \frac{az+b+\alpha\theta}{cz+d+\beta\theta} \;,
\qquad
\theta  \mapsto  
  - \alpha + \beta \frac{az+b}{cz+d} +\frac{\theta}{cz+d} \;,
\label{eq:superproj}
\end{equation}
where $a,b,c,d$ are Grassmann even, $\alpha,\beta$ are 
Grassmann odd and they satisfy
\begin{equation}
ad-bc=1+\alpha\beta\;.
\end{equation}
Therefore there are $N-3$ Grassmann even and 
$N-2$ Grassmann odd moduli parameters. 

In order to define the moduli parameters explicitly, 
we need to define the interaction points on the super
light-cone diagram. 
The definition of the interaction points 
is a little bit complicated 
compared with the bosonic case
\cite{Berkovits:1985ji,Berkovits:1987gp,Aoki:1990yn,Mandelstam1992}.
One can define 
$\tilde{\mathbf{z}}_{I}\;\left(I=1,\cdots,N-2\right)$
which satisfy
\begin{equation}
\partial\rho\left(\tilde{\mathbf{z}}_{I}\right) =  0 \;,
\qquad
\partial D\rho\left(\tilde{\mathbf{z}}_{I}\right)
   =  0 \;.
\label{eq:int-pt-tilde}
\end{equation}
The interaction points $\tilde{\mathbf{z}}_{I}$'s may be considered 
as a straightforward generalization
of the bosonic version, 
but they do not transform covariantly under
the superprojective transformations.
In order to remedy the covariance,
in place of $\tilde{\mathbf{z}}_{I}$
we introduce $\mathbf{z}_{I}$ \cite{Berkovits:1987gp,Aoki:1990yn}
which is defined so that with an appropriately
chosen Grassmann odd parameter $\xi_{I}$, 
\begin{equation}
\hat{\rho}\left(\mathbf{z}\right) 
 \equiv \rho\left(\mathbf{z}\right)
   -\rho\left(\mathbf{z}_{I}\right)
   -\left[ 2 \left( \rho\left(\mathbf{z}\right)
                    -\rho\left(\mathbf{z}_{I}\right)
             \right)
     \right]^{-\frac{1}{4}} 
     \xi \left( \mathbf{z} \right)
     \xi_{I} 
\label{eq:hatrhodef}
\end{equation}
can be expanded as
\begin{equation}
\hat{\rho}\left(\mathbf{z}\right)
  =  \frac{1}{2} \partial^{2} \hat{\rho} \left(\mathbf{z}_{I}\right)
     \left( \mathbf{z}-\mathbf{z}_{I} \right)^{2}
    + \cdots
\label{eq:hatrhoexp}
\end{equation}
for $\mathbf{z}\sim\mathbf{z}_{I}$.
We summarize properties of $\mathbf{z}_{I}$ and $\xi_{I}$
in appendix \ref{sec:int-points}. 
$\xi_{I}$ $\left(I=1,\cdots,N-2\right)$
correspond to the odd moduli of the super Riemann surface
and are determined in eq.(\ref{eq:xiIdef}).
As we see in appendix \ref{sec:int-points},
the above definition of $\mathbf{z}_{I}$
leads to eq.(\ref{eq:zI-another}).
It follows that the interaction points $\mathbf{z}_{I}$
can be considered to be defined as the solutions
to the equations \cite{Mandelstam1992,Aoki:1990yn}
\begin{equation}
\partial \rho (\mathbf{z})
 -\frac{1}{2} 
  \frac{\partial^{2}D\rho D\rho}{\partial^{2}\rho}
    (\mathbf{z})=0\;,
\qquad
\partial D \rho (\mathbf{z})
 - \frac{1}{6} \frac{\partial^{3}\rho D\rho}
                    {\partial^{2} \rho} (\mathbf{z})
 =0~.
\end{equation}
Since these equations transform covariantly
under the superprojective transformations,
the interaction points
$\mathbf{z}_{I}$ transform covariantly as well
and thus have a physical meaning.
As shown in appendix \ref{sec:int-points},
$D\rho(\tilde{\mathbf{z}}_{I})
  \left( =D\rho(\mathbf{z}_{I}) \right)$
and the difference between 
$\mathbf{z}_{I}$ and $\tilde{\mathbf{z}}_{I}$
are proportional to $\xi_{I}$.
This yields
\begin{eqnarray}
\rho\left(\mathbf{z}_{I}\right) 
 & = & \rho\left(\tilde{\mathbf{z}}_{I}\right)
        + \left(\theta_{I}-\tilde{\theta}_{I}\right)
          D\rho\left(\tilde{\mathbf{z}}_{I}\right)
        + \left(\mathbf{z}_{I}-\tilde{\mathbf{z}}_{I}\right)
          \partial\rho\left(\tilde{\mathbf{z}}_{I}\right)
\nonumber \\
 & = & \rho\left(\tilde{\mathbf{z}}_{I}\right)\;.
\label{eq:rhozIeq}
\end{eqnarray}
Therefore the even moduli parameters $\mathcal{T}_{I}$
$\left(I=1,\cdots,N-3\right)$
can be defined as
\begin{eqnarray}
\mathcal{T}_{I} 
& \equiv & \rho\left(\mathbf{z}_{I+1}\right)
           -\rho\left(\mathbf{z}_{I}\right)
\nonumber \\
 & = & \rho\left(\tilde{\mathbf{z}}_{I+1}\right)
        -\rho\left(\tilde{\mathbf{z}}_{I}\right)\;.
\end{eqnarray}

Now we would like to calculate the variation of 
$-\Gamma_{\mathrm{super}}$
when we modify the moduli parameters as 
$\mathcal{T}_{I} \mapsto\mathcal{T}_{I}+\delta\mathcal{T}_{I}$,
$\xi_{I}\to\xi_{I}$.
For this variation, we can calculate 
$\delta\left(-\Gamma_{\mathrm{super}}\right)$
in a way similar to the bosonic case \cite{Baba:2009ns} 
and we obtain
\begin{eqnarray}
\delta\left(-\Gamma_{\mathrm{super}}\right) 
& = & 
  \sum_{I} \delta\mathcal{T}_{I}
    \oint_{C_{I}} \frac{d\mathbf{z}}{2\pi i}
       \frac{\left(-2S\left(\mathbf{z},\boldsymbol{\rho}\right)\right)}
            {\left(D\xi\right)^{2}\left(\mathbf{z}\right)}
\nonumber\\
& = & 
  \sum_{r}\oint_{\mathbf{Z}_{r}}\frac{d\mathbf{z}}{2\pi i}
      \frac{\delta\rho(\mathbf{z}_{I}^{(r)})}
           {\left(D\xi\right)^{2}\left(\mathbf{z}\right)}
      \left(-2S\left(\mathbf{z},\boldsymbol{\rho}\right)\right)
%\nonumber \\
% &  & \quad 
   {}+ \sum_{I} \oint_{\mathbf{z}_{I}} \frac{d\mathbf{z}}{2\pi i}
         \frac{\delta\rho\left(\mathbf{z}_{I}\right)}
              {\left(D\xi\right)^{2}\left(\mathbf{z}\right)}
         \left(-2S\left(\mathbf{z},\boldsymbol{\rho}\right)\right)
\nonumber \\
 &  & \qquad
   {}+ \oint_{\infty}\frac{d\mathbf{z}}{2\pi i}
        \frac{\delta\rho (\mathbf{z}_{I}^{(\infty)})}
             {\left(D\xi\right)^{2} \left(\mathbf{z}\right)}
       \left(-2S\left(\mathbf{z},\boldsymbol{\rho}\right)\right)
\nonumber \\
 &  &  
   {} +\mathrm{c.c.}\;,
\end{eqnarray}
where the integration contour $C_{I}$ lies
between the consecutive interaction points
$\rho (\mathbf{z}_{I+1})$ and $\rho (\mathbf{z}_{I})$
as in the case of bosonic string \cite{Baba:2009ns},
$\mathbf{z}_{I}^{(r)}$ denotes the interaction
point where the $r$-th external string interacts
and $\mathbf{z}_{I}^{(\infty)}$ denotes the interaction point
closest to $\infty$. 
Using the fact that the possible singularities of
$\frac{\delta\rho\left(\mathbf{z}\right)
        - \delta\xi\xi\left(\mathbf{z}\right)}
      {\left(D\xi\right)^{2}\left(\mathbf{z}\right)}
  S\left(\mathbf{z},\boldsymbol{\rho}\right)$
are at $\mathbf{z}=\mathbf{Z}_{r},\mathbf{z}_{I},\infty$,
we can express the variation of $-\Gamma_{\mathrm{super}}$ as
\begin{eqnarray}
\delta\left(-\Gamma_{\mathrm{super}}\right)
 & = & -\sum_{r} \oint_{\mathbf{Z}_{r}}
   \frac{d\mathbf{z}}{2\pi i}
   \frac{\delta \rho \left(\mathbf{z}\right)
         - \delta\rho (\mathbf{z}_{I}^{(r)} )
         - \delta\xi\xi\left(\mathbf{z}\right)}
        {\left(D\xi\right)^{2}\left(\mathbf{z}\right)}
  \left(-2S\left(\mathbf{z},\boldsymbol{\rho}\right)\right)
\nonumber \\
 &  & \quad
   {}-\sum_{I} \oint_{\mathbf{z}_{I}} \frac{d\mathbf{z}}{2\pi i}
       \frac{\delta\rho\left(\mathbf{z}\right)
              -\delta\rho\left(\mathbf{z}_{I}\right)
              -\delta\xi\xi\left(\mathbf{z}\right)}
            {\left(D\xi\right)^{2}\left(\mathbf{z}\right)}
       \left(-2S\left(\mathbf{z},\boldsymbol{\rho}\right)\right)
\nonumber \\
 &  & \quad
   {}-\oint_{\infty}\frac{d\mathbf{z}}{2\pi i}
     \frac{\delta\rho\left(\mathbf{z}\right)
            -\delta \rho (\mathbf{z}_{I}^{(\infty)})
            -\delta\xi\xi\left(\mathbf{z}\right)}
          {\left(D\xi\right)^{2}\left(\mathbf{z}\right)}
     \left(-2S\left(\mathbf{z},\boldsymbol{\rho}\right)\right)
\nonumber \\
 &  & {}+\mathrm{c.c.} \;.
\label{eq:ward}
\end{eqnarray}
We will evaluate $\delta\left(-\Gamma_{\mathrm{super}}\right)$
by performing the contour integrations, 
and integrate it to obtain $-\Gamma_{\mathrm{super}}$. 
We cannot fix the dependence on $\xi_{I}$ and $\alpha_{r}$ 
by this method. 
However,
since the dependence on $\mathcal{T}_{I}$ is known, 
we can take the limit $\mathcal{T}_{I}\to\infty$. 
Imposing the factorization conditions
on $-\Gamma_{\mathrm{super}}$,
the entire $-\Gamma_{\mathrm{super}}$ can be obtained. 

In order to calculate the right hand side of eq.(\ref{eq:ward}),
we need to know the behavior of the integrand for
$\mathbf{z}\sim\mathbf{Z}_{r},\mathbf{z}_{I},\infty$.
The super Schwarzian derivative 
$S\left(\mathbf{z},\boldsymbol{\rho}\right)$
can be written as
\begin{equation}
S\left(\mathbf{z},\boldsymbol{\rho}\right) 
 =  -\frac{1}{4}D\ln\omega (\mathbf{z})
                \partial\ln\omega (\mathbf{z})
    + \frac{1}{2}\partial D\ln\omega (\mathbf{z})\;,
\end{equation}
where
\begin{equation}
\omega (\mathbf{z}) \equiv 
 \left(D\xi\right)^{2} (\mathbf{z})
 =  \partial \rho (\mathbf{z})
    - \frac{\partial D\rho D\rho}{\partial\rho} (\mathbf{z}) \;.
\label{eq:omegadef}
\end{equation}
$\omega$ plays a role similar to $\partial\rho$ in the bosonic case.
In the bosonic case, the fact that $\partial\rho$ can be written
as
\begin{equation}
\partial\rho\left(z\right)  
=  \frac{\left( \sum_{r}\alpha_{r}Z_{r} \right)
          \prod_{I}\left( z-z_{I} \right)}
        {\prod_{s} \left( z-Z_{s} \right)} 
\label{eq:partialrho}
\end{equation}
makes the calculations easier. 
Here we need a similar identity for $\omega$. 

%%%%%%%%%%%%%%%%%%%%%%%%%%%%%
\subsection{$\omega \left(\mathbf{z} \right)$}
In order to obtain an expression like eq.(\ref{eq:partialrho}) 
for $\omega$, 
we need some facts about the polynomials of the supercoordinate
$\mathbf{z}=(z,\theta)$. 

\subsubsection*{Polynomials in superspace}
We restrict ourselves to the situation in which there are
a finite number of Grassmann odd parameters. 
In general,
a Grassmann even number $a$ can be decomposed as
\begin{equation}
a =  a^{\left(0\right)}+a^{\left(2\right)}+a^{\left(4\right)}
    +\cdots \;,
\end{equation}
where $a^{\left(n\right)}$ involves $n$ Grassmann odd parameters.
In the present situation, such an expansion terminates
at a finite order.
It is easy to prove that 
if $a^{\left(0\right)}\neq0$, there exists 
the inverse $a^{-1}$ of $a$. 

Let us consider a Grassmann even superanalytic function 
$f\left(\mathbf{z}\right)$ in superspace.
This can be expressed as
\begin{equation}
f\left(\mathbf{z}\right)
=  f_{0}\left(z\right) +\theta f_{1}\left(z\right) \;.
\label{eq:polynomial}
\end{equation}
In the following, let us assume that $f_{0}\left(z\right)$
and $f_{1}\left(z\right)$ are $N$-th order polynomials of $z$. 
The polynomial $f_{0}\left(z\right)$
can also be decomposed as 
\begin{equation}
f_{0}\left(z\right)  
=  f_{0}^{(0)} (z)+f_{0}^{(2)} (z) +\cdots \;,
\end{equation}
where the coefficients of $f_{0}^{(n)}$ involves $n$
Grassmann odd parameters. 
$f^{(0)}_{0}(z)$ is referred to as the body part
of $f(\mathbf{z})$.

If there exists 
$\mathbf{z}_{i}$=$\left(z_{i},\theta_{i}\right)$
such that
\begin{equation}
f\left(\mathbf{z}_{i}\right) 
 =  Df\left(\mathbf{z}_{i}\right)=0 \;,
\label{eq:zI}
\end{equation}
by Taylor expanding 
$f\left(\mathbf{z}\right)$ around $\mathbf{z}=\mathbf{z}_{i}$,
one can show that 
\begin{equation}
f\left(\mathbf{z}\right) 
 =  \left(\mathbf{z}-\mathbf{z}_{i}\right)
     \tilde{f}\left(\mathbf{z}\right)\;,
\qquad
 \tilde{f} (\mathbf{z})
    = \tilde{f}_{0}\left(z\right)
            +\theta\tilde{f}_{1}\left(z\right)\;,
\label{eq:ftilde}
\end{equation}
where $\tilde{f}_{0}\left(z\right)$ and 
$\tilde{f}_{1}\left(z\right)$ are
polynomials of $z$ whose orders are at most $N-1$. 
Suppose that there exists 
$\mathbf{z}_{j}=\left(z_{j},\theta_{j}\right)$ 
for this
$f\left(\mathbf{z}\right)$ such that 
$z_{j}^{(0)} \neq z_{i}^{(0)}$ and
\begin{equation}
f\left(\mathbf{z}_{j}\right) 
  =  Df\left(\mathbf{z}_{j}\right) =0 \;.
\label{eq:zJ}
\end{equation}
Substituting eq.(\ref{eq:ftilde}) into eq.(\ref{eq:zJ}),
one obtains
\begin{eqnarray}
\left(\mathbf{z}_{j}-\mathbf{z}_{i}\right)
  \tilde{f}\left(\mathbf{z}_{j}\right) & = & 0 \;,
\nonumber \\
\left(\theta_{j}-\theta_{i}\right) 
   \tilde{f}\left(\mathbf{z}_{j}\right)
  + \left( \mathbf{z}_{j} - \mathbf{z}_{i} \right)
      D\tilde{f}\left(\mathbf{z}_{j}\right) & = & 0 \;.
\end{eqnarray}
Since $\mathbf{z}_{j}-\mathbf{z}_{i}$ has the inverse, 
one can show that
$\tilde{f}\left(\mathbf{z}_{j}\right)
  =D\tilde{f}\left(\mathbf{z}_{j}\right)=0$
and thus $f\left(\mathbf{z}\right)$ can be further factorized as
\begin{equation}
f\left(\mathbf{z}\right) 
 =  \left(\mathbf{z}-\mathbf{z}_{i}\right)
    \left(\mathbf{z}-\mathbf{z}_{j}\right)
      \tilde{\tilde{f}}\left(\mathbf{z}\right) \;.
\end{equation}
If one can go on like this, 
one can show that the function $f\left(\mathbf{z}\right)$
can be factorized as the usual polynomials. 

However, this is not always the case. 
For example, for 
\begin{equation}
f\left(\mathbf{z}\right) = z^{2}+a\;,
\end{equation}
with $a^{\left(0\right)}=0$, 
one cannot find a solution to eq.(\ref{eq:zI})
and $f\left(\mathbf{z}\right)$ cannot be factorized. 
The problem about this kind of polynomials is 
that the body part has a double root. 
Indeed one can show that if $f_{0}^{\left(0\right)}\left(z\right)$
can be factorized as
\begin{equation}
f_{0}^{\left(0\right)}\left(z\right) 
=  a \prod_{i}\left(z-z_{i}^{\left(0\right)}\right) \;,
\end{equation}
with $z_{i}^{(0)}\neq z_{j}^{(0)}$ for $i\neq j$, 
$f\left(\mathbf{z}\right)$ can be factorized. 
One can find $z_{i}$ satisfying $f_{0}\left(z_{i}\right)=0$ 
by expanding
\begin{equation}
z_{i}  =  z_{i}^{\left(0\right)}+z_{i}^{\left(2\right)}
          +z_{i}^{\left(4\right)}+ \cdots \;.
\end{equation}
$z_{i}^{\left(2n\right)}$ can be obtained order by order 
by solving equations of the form
\begin{equation}
z_{i}^{\left(2n\right)}
  \prod_{j\neq i} \left(z_{i}^{\left(0\right)}
   -z_{j}^{\left(0\right)}\right)
   + \left(\mbox{contributions from lower order}\right) 
 =  0\;.
\end{equation}
Then one can show that 
$\mathbf{z}_{i}
  =\left( z_{i}, -\frac{f_{1}\left(z_{i}\right)}
                       {\partial f_{0}\left(z_{i}\right)}\right)$
satisfies 
$f\left(\mathbf{z}_{i}\right)
  =Df\left(\mathbf{z}_{i}\right)=0$.

%%%%%%%%%%%%%%%%%%%%%%%%%%%%%%%%%%%%%%%%%%%%
\subsubsection*{Factorization of $\omega (\mathbf{z})$}

Using above facts, it is easy to show
\begin{equation}
\partial\rho\left(\mathbf{z}\right)
=  \sum_{r}\frac{\alpha_{r}}{\mathbf{z}-\mathbf{Z}_{r}}
=
\frac{\left( \sum_{r}  \alpha_{r}Z_{r}
                    + \theta \sum_{r}  \alpha_{r} \Theta_{r}
              \right)
                \prod_{I}
                 \left(\mathbf{z}-\tilde{\mathbf{z}}_{I}\right)}
            {\prod_{r}\left(\mathbf{z}-\mathbf{Z}_{r}\right)} \;,
\end{equation}
assuming that the body part $z_{I}^{\left(0\right)}$
satisfies $z_{I}^{\left(0\right)}\neq z_{J}^{\left(0\right)}$ 
for $I\neq J$. One can also show 
\begin{equation}
-\frac{\partial D\rho D\rho}{\partial\rho} (\mathbf{z})
 =  \frac{\sum_{r} 
           \left( \alpha_{r}\left(\theta-\Theta_{r}\right)
                  \prod_{s\neq r}
                     \left(\mathbf{z}-\mathbf{Z}_{s}\right)^{2}
           \right)
          \sum_{r}
            \left( \alpha_{r}\left(\theta-\Theta_{r}\right)
                   \prod_{s\neq r}
                      \left(\mathbf{z}-\mathbf{Z}_{s}\right)
            \right)}
          {\left( \sum_{r} \alpha_{r}Z_{r}
                  + \theta \sum_{r} \alpha_{r}\Theta_{r}
             \right)
            \prod_{r}\left(\mathbf{z}-\mathbf{Z}_{r}\right)^{2}
            \prod_{I}\left(\mathbf{z}-\tilde{\mathbf{z}}_{I}\right)}
\;.
\end{equation}
It is easy to see that the numerator is factorized as 
$$
F\left(\mathbf{z}\right)
 \prod_{r}\left(\mathbf{z}-\mathbf{Z}_{r}\right) \;,
$$
where $F\left(\mathbf{z}\right)$ is a polynomial of order $2N-4$
whose coefficients involve at least two Grassmann odd parameters. 
Therefore $\omega$ can be written as
\begin{equation}
\omega 
= \frac{\left( \sum_{r} \alpha_{r}Z_{r}
                      + \theta \sum_{r} \alpha_{r} \Theta_{r}
        \right)^{2}
        \prod_{I}
          \left(\mathbf{z}-\tilde{\mathbf{z}}_{I}\right)^{2}
        + F\left(\mathbf{z}\right)}
       {\left( \sum_{r}  \alpha_{r}Z_{r}
                 + \theta \sum_{r} \alpha_{r}\Theta_{r}
         \right)
        \prod_{r} \left(\mathbf{z}-\mathbf{Z}_{r}\right)
        \prod_{I}\left(\mathbf{z}-\tilde{\mathbf{z}}_{I}\right)}
\;.
\end{equation}
It is obvious that the body part of the numerator has double roots. 

In order to get a factorized form of $\omega$, 
let us ``regularize'' it \cite{Mandelstam1992,Aoki:1990yn}
by introducing
\begin{equation}
\omega_{\epsilon} 
 \equiv 
 \frac{\left(\sum_{r} \alpha_{r}Z_{r}
             +\theta \sum_{r} \alpha_{r}\Theta_{r}
        \right)^{2}
        \prod_{I}\left(\mathbf{z}-\tilde{\mathbf{z}}_{I}\right)^{2}
        + F \left(\mathbf{z}\right)
        - \epsilon^{2}}
      {\left(\sum_{r} \alpha_{r}Z_{r}
             +\theta\sum_{r} \alpha_{r}\Theta_{r}
       \right)
       \prod_{r}\left(\mathbf{z}-\mathbf{Z}_{r}\right)
       \prod_{I}\left(\mathbf{z}-\tilde{\mathbf{z}}_{I}\right)}\;,
\end{equation}
where $\epsilon ~(|\epsilon |\ll 1)$ is a complex number. 
$\omega_{\epsilon}$ can be factorized as
\begin{equation}
\omega_{\epsilon} 
 =  A \frac{\prod_{I}\left[
             \left(\mathbf{z}-\mathbf{z}_{I+}\right)
             \left(\mathbf{z}-\mathbf{z}_{I-}\right)
            \right]}
           {\prod_{r}\left(\mathbf{z}-\mathbf{Z}_{r}\right)
            \prod_{I}\left(\mathbf{z}-\tilde{\mathbf{z}}_{I}
                     \right)} \;,
\label{eq:omegaep}
\end{equation}
where 
\begin{equation}
A \equiv  \sum_{r} \alpha_{r}Z_{r}
 -\frac{\sum_{r} \alpha_{r}\Theta_{r}
        \sum_{r} \alpha_{r}\Theta_{r}Z_{r}}
       {\sum_{r} \alpha_{r}Z_{r}}\;.
\end{equation}
Let us define $B,\gamma$ as 
\begin{eqnarray}
&& B^{2} = - g (\tilde{\mathbf{z}}_{I}) \;,
   \qquad
   \gamma= D g(\tilde{\mathbf{z}}_{I})\;,
\nonumber\\
&& g(\mathbf{z})
   = - \frac{1}
            {\left( \sum_{r} \alpha_{r} Z_{r}
                      + \theta \sum_{r} \alpha_{r}\Theta_{r}
                \right)
             \prod_{r}\left( \mathbf{z}-\mathbf{Z}_{r} \right)
             \prod_{J\neq I}
                \left( \mathbf{z}-\tilde{\mathbf{z}}_{J}
                      \right)}\;,
\end{eqnarray}
so that
\begin{equation}
\lim_{\mathbf{z}\rightarrow \tilde{\mathbf{z}}_{I}}
 \left[ \left(\mathbf{z}-\tilde{\mathbf{z}}_{I}\right)
        \omega_{\epsilon}\left(\mathbf{z}\right)
 \right]
=  -\epsilon^{2}B^{2} \;,
\qquad
\lim_{\mathbf{z}\rightarrow \tilde{\mathbf{z}}_{I}}
  D \left[ \left(\mathbf{z}-\tilde{\mathbf{z}}_{I}\right)
           \omega_{\epsilon}\left(\mathbf{z}\right)
    \right]
= -D\rho\left(\tilde{\mathbf{z}}_{I}\right)
  + \epsilon^{2}\gamma \;.
\end{equation}
Using these variables, we obtain
for  $|\epsilon |\ll 1$
\begin{eqnarray}
\mathbf{z}_{I\pm}-\tilde{\mathbf{z}}_{I} 
 & \sim & \pm \frac{\epsilon B}
                   {\left(\partial^{2}\rho\right)^{\frac{1}{2}}
                       \left(\tilde{\mathbf{z}}_{I}\right)}
          + \frac{\partial^{2} D\rho D\rho}
                 {2\left(\partial^{2}\rho\right)^{2}}
               \left(\tilde{\mathbf{z}}_{I}\right)\;,
\nonumber \\
\theta_{I\pm}-\tilde{\theta}_{I}
 & \sim & \pm \frac{1}{2\epsilon B}
            \frac{D\rho}
                 {\left(\partial^{2}\rho\right)^{\frac{1}{2}}}
               \left(\tilde{\mathbf{z}}_{I}\right)
          \mp \frac{3}{4} \epsilon B 
            \frac{\partial^{2}D\rho}
                 {\left(\partial^{2}\rho\right)^{2}}
              \left(\tilde{\mathbf{z}}_{I}\right)
          \mp \frac{\epsilon\gamma}
                   {2B\left(\partial^{2}\rho\right)^{\frac{1}{2}}
              \left(\tilde{\mathbf{z}}_{I}\right)} \;.
\end{eqnarray}
Hence, taking the limit $\epsilon\to0$ in eq.(\ref{eq:omegaep}), we obtain
\begin{equation}
\omega  = \lim_{\epsilon\to0}\omega_{\epsilon}
 =  A \prod_{I}\left[
         \left( \mathbf{z}-\tilde{\mathbf{z}}_{I} \right)
         - \frac{\partial^{2}D\rho D\rho}
                {\left(\partial^{2}\rho\right)^{2}}
              \left(\tilde{\mathbf{z}}_{I}\right)
         - \frac{\theta-\tilde{\theta_{I}}}
                {\mathbf{z}-\tilde{\mathbf{z}}_{I}}
           \frac{D\rho}{\partial^{2}\rho}
             \left(\tilde{\mathbf{z}}_{I}\right)
       \right]
     \frac{1}{\prod_{r}\left(\mathbf{z}-\mathbf{Z}_{r}\right)}\;.
\label{eq:omegafact}
\end{equation}

%%%%%%%%%%%%%%%%%%%%%%%%%%%%%%%%%%%%%%%%%
\subsection{Calculation of the contour integrals}

With the form of $\omega$ in eq.(\ref{eq:omegafact}), 
one can calculate the contour integrals 
on the right hand side of eq.(\ref{eq:ward}). 

For $\mathbf{z}\sim\mathbf{Z}_{r}$, the super Schwarzian derivative
behaves as 
\begin{equation}
S\left(\mathbf{z},\boldsymbol{\rho}\right)
  \sim  \frac{1}{4}
         \frac{\theta-\Theta_{r}}
              {\left(\mathbf{z}-\mathbf{Z}_{r}\right)^{2}}
       + \frac{1}{\mathbf{z}-\mathbf{Z}_{r}}
         \frac{1}{4}
         D\ln f_{r}\left(\mathbf{Z}_{r}\right)
       + \frac{\theta-\Theta_{r}}{\mathbf{z}-\mathbf{Z}_{r}}
         \frac{1}{2}
         \partial \ln f_{r} \left(\mathbf{Z}_{r}\right)\;,
\end{equation}
where 
\begin{equation}
\ln f_{r}\left(\mathbf{z}\right)
 = \ln A
     + \sum_{I} \ln\left[ 
          \left( \mathbf{z}-\tilde{\mathbf{z}}_{I} \right)
          - \frac{\partial^{2}D\rho D\rho}
                 {\left(\partial^{2}\rho\right)^{2}}
              \left(\tilde{\mathbf{z}}_{I}\right)
          - \frac{\theta-\tilde{\theta_{I}}}
                 {\mathbf{z}-\tilde{\mathbf{z}}_{I}}
            \frac{D\rho}{\partial^{2}\rho}
              \left(\tilde{\mathbf{z}}_{I}\right)
                      \right]
%\nonumber \\
% &  & \hspace{1cm}
       {} -\sum_{s\neq r}\ln\left(\mathbf{z}-\mathbf{Z}_{s}\right)\;.
\end{equation}
Thus it is straightforward to obtain
\begin{eqnarray}
\lefteqn{
 \oint_{\mathbf{Z}_{r}} \frac{d\mathbf{z}}{2\pi i}
    \frac{\delta \rho \left(\mathbf{z}\right)
          - \delta \rho (\mathbf{z}_{I}^{(r)} )
          -\delta \xi \xi \left(\mathbf{z}\right)}
         {\left(D\xi\right)^{2}\left(\mathbf{z}\right)}
    2S\left(\mathbf{z},\boldsymbol{\rho}\right)
} \nonumber\\
 &  & = {}- \left( \delta Z_{r} - \delta\Theta_{r}\Theta_{r} \right)
            \partial\ln f_{r}\left(\mathbf{Z}_{r}\right)
        - \delta \Theta_{r} D\ln f_{r}\left(\mathbf{Z}_{r}\right)
        - \frac{1}{2} \delta \bar{N}_{00}^{rr}\;,
\label{eq:integralZr}
\end{eqnarray}
where
\begin{equation}
\bar{N}_{00}^{rr} 
=  \frac{\rho (\tilde{\mathbf{z}}_{I}^{(r)})}
        {\alpha_{r}}
   - \sum_{s\neq r} \frac{\alpha_{s}}{\alpha_{r}}
         \ln\left(\mathbf{Z}_{r}-\mathbf{Z}_{s}\right)\;.
\end{equation}

For $\mathbf{z}\sim\infty$, it is easy to see that
\begin{equation}
S \left(\mathbf{z},\boldsymbol{\rho}\right) 
 \sim  0\cdot\frac{\theta}{z^{3}}
       +\mathcal{O}\left(z^{-3}\right)\;,
\qquad
 \frac{\delta\rho\left(\mathbf{z}\right)
         -\delta\rho\left(\mathbf{z}_{I}\right)
         -\delta\xi\xi\left(\mathbf{z}\right)}
        {\left(D\xi\right)^{2}\left(\mathbf{z}\right)} 
 \sim  \mathcal{O}\left(z^{2}\right)\;,
\end{equation}
and we obtain
\begin{equation}
\oint_{\infty} \frac{d\mathbf{z}}{2\pi i}
  \frac{\delta \rho \left(\mathbf{z}\right)
        -\delta\rho (\mathbf{z}_{I}^{(\infty)})
        -\delta\xi\xi\left(\mathbf{z}\right)}
       {\left(D\xi\right)^{2}\left(\mathbf{z}\right)}
  S\left(\mathbf{z},\boldsymbol{\rho}\right) =  0\;.
\label{eq:integralinfty}
\end{equation}

The most difficult is the integral around $\mathbf{z}_{I}$. From
eq.(\ref{eq:omegafact}) one can derive 
\begin{eqnarray}
S\left(\mathbf{z},\boldsymbol{\rho}\right) 
& \sim & 
  \frac{1}{\left(\mathbf{z}-\tilde{\mathbf{z}}_{I}\right)^{3}}
    \frac{5}{4}\frac{D\rho}{\partial^{2}\rho}
        \left(\tilde{\mathbf{z}}_{I}\right)
\nonumber \\
 &  &
    {}+\frac{\theta-\tilde{\theta}_{I}}
            {\left(\mathbf{z}-\tilde{\mathbf{z}}_{I}\right)^{3}}
     \left(-\frac{3}{2}\frac{\partial^{2}D\rho D\rho}
                            {\left(\partial^{2}\rho\right)^{2}}
                          \left(\tilde{\mathbf{z}}_{I}\right)
           -\frac{1}{2} \frac{D\rho}{\partial^{2}\rho}
                          \left(\tilde{\mathbf{z}}_{I}\right)
             D\ln f_{I}\left(\tilde{\mathbf{z}}_{I}\right)
     \right)
\nonumber \\
 &  &
   {}+ \frac{1}{\left(\mathbf{z}-\tilde{\mathbf{z}}_{I}\right)^{2}}
       \left(\frac{1}{4}
              \frac{D\rho}{\partial^{2}\rho}
                \left(\tilde{\mathbf{z}}_{I}\right)
              \partial\ln f_{I}\left(\tilde{\mathbf{z}}_{I}\right)
            - \frac{1}{4}
              \frac{\partial^{2}D\rho D\rho}
                   {\left(\partial^{2}\rho\right)^{2}}
                \left(\tilde{\mathbf{z}}_{I}\right)
              D\ln f_{I}\left(\tilde{\mathbf{z}}_{I}\right)
        \right)
\nonumber \\
 &  &
  {}+\frac{\theta-\tilde{\theta}_{I}}
          {\left(\mathbf{z}-\tilde{\mathbf{z}}_{I}\right)^{2}}
     \left(-\frac{3}{4}
             \frac{D\rho}{\partial^{2}\rho}
                \left(\tilde{\mathbf{z}}_{I}\right)
             \partial D\ln f_{I}
                \left(\tilde{\mathbf{z}}_{I}\right)
     \vphantom{\frac{\partial^{2}D\rho D\rho}
                    {\left(\partial^{2}\rho\right)^{2}}}
     \right.\nonumber \\
 &  & 
    \hphantom{+\frac{\theta-\tilde{\theta}_{I}}
                    {\left(\mathbf{z}-\tilde{\mathbf{z}}_{I}
                     \right)^{2}}
             }
     \ \left.{}-\frac{1}{2}
       \frac{\partial^{2}D\rho D\rho}
            {\left(\partial^{2}\rho\right)^{2}}
           \left(\tilde{\mathbf{z}}_{I}\right)
       \partial\ln f_{I}\left(\tilde{\mathbf{z}}_{I}\right)
      - \frac{3}{4}\right)
\nonumber \\
 &  &
  {}+\frac{1}{\mathbf{z}-\tilde{\mathbf{z}}_{I}}
    \left( -\frac{1}{4}D\ln f_{I}
              \left(\tilde{\mathbf{z}}_{I}\right)
           -\frac{1}{4}
             \frac{\partial^{2}D\rho D\rho}
                  {\left(\partial^{2}\rho\right)^{2}}
               \left(\tilde{\mathbf{z}}_{I}\right)
            \partial D\ln f_{I}\left(\tilde{\mathbf{z}}_{I}\right)
    \right.\nonumber \\
 &  & 
   \hphantom{+\frac{1}{\mathbf{z}-\tilde{\mathbf{z}}_{I}}}
   \ \left.{}+\frac{1}{4}
      \frac{D\rho}{\partial^{2}\rho}
         \left(\tilde{\mathbf{z}}_{I}\right)
      \partial^{2} \ln f_{I}\left(\tilde{\mathbf{z}}_{I}\right)
    \vphantom{\frac{\partial^{2}D\rho D\rho}
                   {\left(\partial^{2}\rho\right)^{2}}
                \left(\tilde{\mathbf{z}}_{I}\right)
                \partial D\ln f_{I}\left(\tilde{\mathbf{z}}_{I}\right)
              }
     \right) \nonumber \\
 &  & 
   {}+\frac{\theta-\tilde{\theta}_{I}}
           {\mathbf{z}-\tilde{\mathbf{z}}_{I}}
      \left( -\frac{1}{2} \partial\ln f_{I}
                  \left(\tilde{\mathbf{z}}_{I}\right)
             -\frac{1}{2}
                 \frac{\partial^{2}D\rho D\rho}
                      {\left(\partial^{2}\rho\right)^{2}}
                        \left(\tilde{\mathbf{z}}_{I}\right)
                 \partial^{2} \ln f_{I}
                    \left(\tilde{\mathbf{z}}_{I}\right)
      \right.\nonumber \\
 &  &
   \hphantom{+\frac{\theta-\tilde{\theta}_{I}}
                   {\mathbf{z}-\tilde{\mathbf{z}}_{I}}
             }
   \ \left. {}-\frac{1}{2}
       \frac{D\rho}{\partial^{2}\rho}
            \left(\tilde{\mathbf{z}}_{I}\right)
        \partial^{2} D \ln f_{I}
            \left(\tilde{\mathbf{z}}_{I}\right)
      \right)\;,
\label{eq:superschzI}
\end{eqnarray}
where
$\ln f_{I} (\mathbf{z})$ is defined by
\begin{eqnarray}
\ln f_{I}\left(\mathbf{z}\right) 
& = &
 \ln A 
 +\sum_{J\neq I} \ln \left[
     \left( \mathbf{z}-\tilde{\mathbf{z}}_{J} \right)
     - \frac{\partial^{2}D\rho D\rho}
            {\left(\partial^{2}\rho\right)^{2}}
              \left(\tilde{\mathbf{z}}_{J}\right)
     - \frac{\theta - \tilde{\theta_{J}}}
            {\mathbf{z}-\tilde{\mathbf{z}}_{J}}
       \frac{D\rho}{\partial^{2}\rho}
          \left(\tilde{\mathbf{z}}_{J}\right) \right]
\nonumber \\
 &  & \hspace{1cm}
   {} -\sum_{r}\ln\left(\mathbf{z}-\mathbf{Z}_{r}\right)\;,
\label{eq:fI-2}
\end{eqnarray}
and thus
\begin{equation}
\omega (\mathbf{z})
  = \left[ \left( \mathbf{z}- \tilde{\mathbf{z}}_{I} \right)
           - \frac{\partial^{2} D \rho D\rho}
                  {(\partial^{2} \rho)^{2}}
               \left(\tilde{\mathbf{z}}_{I} \right)
          - \frac{\theta - \tilde{\theta}_{I}}
                 {\mathbf{z} - \tilde{\mathbf{z}}_{I}}
            \frac{D\rho}{\partial^{2} \rho}
               \left( \tilde{\mathbf{z}}_{I} \right)
    \right]
    f_{I} (\mathbf{z})~.
\label{eq:fI-1}
\end{equation}

In order to see the behavior of 
$\frac{\delta\rho\left(\mathbf{z}\right)
        -\delta\rho\left(\mathbf{z}_{I}\right)
        -\delta\xi\xi\left(\mathbf{z}\right)}
      {\left(D\xi\right)^{2}\left(\mathbf{z}\right)}$
for $\mathbf{z}\sim\mathbf{z}_{I}$, 
it is convenient to introduce
$w(\mathbf{z})$ and $\eta(\mathbf{z})$ such that 
\begin{eqnarray}
\rho\left(\mathbf{z}\right)-\rho\left(\mathbf{z}_{I}\right)
 & = & \frac{1}{2} \left( w(\mathbf{z}) \right)^{2}
         +\eta (\mathbf{z})\xi_{I}\;,
\nonumber \\
\xi\left(\mathbf{z}\right) 
& = & \left( w (\mathbf{z}) \right)^{\frac{1}{2}}\eta (\mathbf{z})
      +\left( w(\mathbf{z}) \right)^{-\frac{1}{2}}\xi_{I}\;,
\label{eq:wdef}
\end{eqnarray}
where $\xi_{I}$ is the Grassmann odd parameter
introduced in eq.(\ref{eq:hatrhodef}).
The map
$\boldsymbol{\rho}=\left( \rho,\xi \right)
  \mapsto \mathbf{w} \equiv \left(w, \eta \right)$
is superconformal, and 
\begin{equation}
\frac{\delta\rho\left(\mathbf{z}\right)
      -\delta\rho\left(\mathbf{z}_{I}\right)
      -\delta\xi\xi\left(\mathbf{z}\right)}
     {\left(D\xi\right)^{2}\left(\mathbf{z}\right)} 
=  \frac{\delta w-\delta\eta\eta}
        {\left(D\eta\right)^{2}} \left(\mathbf{z}\right)
\label{eq:deltarhozI}
\end{equation}
for a variation satisfying $\delta\xi_{I}=0$. 
By using eqs.(\ref{eq:superschzI}) and (\ref{eq:deltarhozI})
and relations given in appendix \ref{sec:int-points},
it is now straightforward to evaluate the contour integral,
\begin{eqnarray}
\lefteqn{
  \oint_{\mathbf{z}_{I}} \frac{d\mathbf{z}}{2\pi i}
     \frac{\delta\rho\left(\mathbf{z}\right)
            -\delta\rho\left(\mathbf{z}_{I}\right)
            -\delta\xi\xi\left(\mathbf{z}\right)}
          {\left(D\xi\right)^{2}\left(\mathbf{z}\right)}
    2 S\left(\mathbf{z},\boldsymbol{\rho}\right)
}
\nonumber \\
 &  & = {}-\left( \delta\tilde{z}_{I}
                  -\delta\tilde{\theta}_{I}\tilde{\theta}_{I}
           \right)
   \partial F_{I} \left(\tilde{\mathbf{z}}_{I}\right)
  -\delta\tilde{\theta}_{I}
  DF_{I}\left(\tilde{\mathbf{z}}_{I}\right)
\nonumber \\
 &  & \hphantom{=}
  {}+ \delta \left( 
          \frac{\partial^{2}D\rho D\rho}
               {\left(\partial^{2}\rho\right)^{2}}
            \left(\tilde{\mathbf{z}}_{I}\right)
         \right)
      \partial\ln f_{I} \left(\tilde{\mathbf{z}}_{I}\right)
    + \delta \left(\frac{D\rho}{\partial^{2}\rho}
                    \left(\tilde{\mathbf{z}}_{I}\right)
             \right)
        \partial D\ln f_{I} \left(\tilde{\mathbf{z}}_{I}\right)
\nonumber \\
 &  & \hphantom{=}
   {}-\frac{3}{4} \delta \left[
       \ln\left(\partial^{2}\rho
                -\frac{13}{9}
                 \frac{\partial^{3}D\rho D\rho}{\partial^{2}\rho}
                + \frac{8}{3}
                  \frac{\partial^{3}\rho\partial^{2}D\rho D\rho}
                       {\left(\partial^{2}\rho\right)^{2}}
            \right) \left(\tilde{\mathbf{z}}_{I}\right)
     \right]\;,
\label{eq:integralzI}
\end{eqnarray}
where
\begin{equation}
F_{I}\left(\mathbf{z}\right) 
\equiv 
  {}-\ln f_{I}\left(\mathbf{z}\right)
    -\frac{\partial^{2}D\rho D\rho}
          {\left(\partial^{2}\rho\right)^{2}}
       \left(\tilde{\mathbf{z}}_{I}\right)
     \partial \ln f_{I} \left(\mathbf{z}\right)
   - \frac{D\rho}{\partial^{2}\rho}
           \left(\tilde{\mathbf{z}}_{I}\right)
      \partial D\ln f_{I}\left(\mathbf{z}\right)\;.
\label{eq:defFI}
\end{equation}

Now we introduce $-W_{\mathrm{super}}$ defined as
\begin{eqnarray}
-W_{\mathrm{super}}
 & \equiv &
   \sum_{r>s} \ln\left(\mathbf{Z}_{r}-\mathbf{Z}_{s}\right)
  + \sum_{I>J} P_{I} P_{J} 
       \ln \left(\tilde{\mathbf{z}}_{I}
                  -\tilde{\mathbf{z}}_{J}\right)
  - \sum_{r} \sum_{I} P_{I}
        \ln \left(\mathbf{Z}_{r}-\tilde{\mathbf{z}}_{I}\right),
\nonumber \\
P_{I}
 & \equiv &
   1 + \frac{\partial^{2}D\rho D\rho}
            {\left(\partial^{2}\rho\right)^{2}}
               \left(\tilde{\mathbf{z}}_{I}\right)
        \tilde{\partial}_{I}
     + \frac{D\rho}{\partial^{2}\rho}
         \left(\tilde{\mathbf{z}}_{I}\right)
       \tilde{\partial}_{I}\tilde{D}_{I}\;,
\label{eq:W-PI}
\end{eqnarray}
where $\tilde{\partial}_{I} = \partial_{\tilde{z}_{I}}$
and 
$\tilde{D}_{I} = \partial_{\tilde{\theta}_{I}}
                 + \tilde{\theta}_{I} \partial_{\tilde{z}_{I}}$.
It satisfies
\begin{eqnarray}
\delta\left(-W_{\mathrm{super}}\right) 
& = & \sum_{r} \left[
       -\left(
        \delta Z_{r}-\delta\Theta_{r}\Theta_{r}\right)
           \partial\ln f_{r}\left(\mathbf{Z}_{r}\right)
         - \delta \Theta_{r} D\ln f_{r}
                      \left(\mathbf{Z}_{r}
        \right)   \right]
 \nonumber \\
 &  & 
   {}+\sum_{I}\left[
        -\left( \delta\tilde{z}_{I}
                 -\delta\tilde{\theta}_{I}\tilde{\theta}_{I}
          \right)
           \partial F_{I} \left(\tilde{\mathbf{z}}_{I}\right)
        -\delta\tilde{\theta}_{I}
           DF_{I}\left(\tilde{\mathbf{z}}_{I}\right)
    \vphantom{\delta\left(\frac{\partial^{2}D\rho D\rho}
                               {\left(\partial^{2}\rho\right)^{2}}
                               \left(\mathbf{z}_{I}\right)\right)
              }\right.
   \nonumber \\
 &  & \hphantom{\sum_{I}}
   \left. \  {}+\delta
          \left( \frac{\partial^{2}D\rho D\rho}
                      {\left(\partial^{2}\rho\right)^{2}}
                     \left(\tilde{\mathbf{z}}_{I}\right)\right)
          \partial \ln f_{I} \left(\tilde{\mathbf{z}}_{I}\right)
           + \delta \left(\frac{D\rho}
                               {\partial^{2}\rho}
                           \left(\tilde{\mathbf{z}}_{I}\right)
                     \right)
              \partial D\ln f_{I}
                  \left(\tilde{\mathbf{z}}_{I}\right)
    \right].~~~~~~
\label{eq:deltaW}
\end{eqnarray}
Putting eqs.(\ref{eq:integralZr}), (\ref{eq:integralinfty}),
(\ref{eq:integralzI}) and (\ref{eq:deltaW}) together,
we finally obtain 
\begin{eqnarray}
\delta\left(-\Gamma_{\mathrm{super}}\right) 
& = & \delta \left[ -W_{\mathrm{super}}
                    -\frac{1}{2}\sum_{r}\bar{N}_{00}^{rr}
     \vphantom{\frac{13}{9}\frac{\partial^{3}D\rho D\rho}
                               {\partial^{2}\rho}
               } \right.
\nonumber \\
 &  & \left. \quad {}-\frac{3}{4} \sum_{I}
       \ln
        \left( \partial^{2}\rho
               -\frac{13}{9}\frac{\partial^{3}D\rho D\rho}
                                 {\partial^{2}\rho}
               + \frac{8}{3} 
                 \frac{\partial^{3}\rho\partial^{2}D\rho D\rho}
                      {\left(\partial^{2}\rho\right)^{2}}
        \right) \left(\tilde{\mathbf{z}}_{I}\right)
      \right] \nonumber \\
 &  & {} +\mbox{c.c.}~.
\label{eq:Gammasuper}
\end{eqnarray}
Since we are dealing with the variation 
satisfying $\delta\xi_{I}=0$,
$\Gamma_{\mathrm{super}}$ is given up to $\xi_{I}$ dependent terms.
In the next subsection, we will fix the $\xi_{I}$ dependent terms
by checking the factorization properties 
and obtain 
\begin{eqnarray}
-\Gamma_{\mathrm{super}}
 & = & -W_{\mathrm{super}}
       -\frac{1}{2}\sum_{r}\bar{N}_{00}^{rr}
\nonumber \\
 &  & \ {} -\frac{3}{4}\sum_{I}
      \ln\left( \partial^{2}\rho
                -\frac{13}{9}
                 \frac{\partial^{3}D\rho D\rho}{\partial^{2}\rho}
                +\frac{8}{3}
                 \frac{\partial^{3}\rho\partial^{2}D\rho D\rho}
                      {\left(\partial^{2}\rho\right)^{2}}
           \right)
        \left(\tilde{\mathbf{z}}_{I}\right)
\nonumber \\
 &  & {}+\mbox{c.c.}~.
\label{eq:Gammasuper2}
\end{eqnarray}

%%%%%%%%%%%%%%%%%%%%%%%%%%%%%%%%%%%%
\subsection{Properties of $\Gamma_{\mathrm{super}}$}

We expect that $\Gamma_{\mathrm{super}}$ 
is invariant under the superprojective
transformation (\ref{eq:superproj}). 
It is easy to see that the expression
on the right hand side of eq.(\ref{eq:ward}) 
is superprojective invariant.
After tedious but straightforward calculations, 
it is possible to check that the right hand side of 
eq.(\ref{eq:Gammasuper2}) is invariant
under the superprojective transformation. 

When all the $\Theta_{r}$'s vanish, 
$\Gamma_{\mathrm{super}}
  \left[\ln\left(\left(D\xi\right)^{2}
           \left(\bar{D}\bar{\xi}\right)^{2}\right)\right]$
should be proportional to the bosonic counterpart 
$\Gamma$ in Ref.~\cite{Baba:2009ns}.
It is easy to see 
\begin{equation}
\left.\Gamma_{\mathrm{super}}
        \left[\ln\left(\left(D\xi\right)^{2}
       \left(\bar{D}\bar{\xi}\right)^{2}\right)\right]
\right|_{\Theta_{r}=0}
 =  \left.\frac{1}{2}
       \Gamma
         \left[\ln\left(\partial\rho\bar{\partial}\bar{\rho}\right)
         \right]\right|_{\Theta_{r}=0}\; .
\end{equation}

%%%%%%%%%%%%%%%%%%%%%%%%%%
\begin{figure}
\begin{centering}
\includegraphics[scale=0.4]{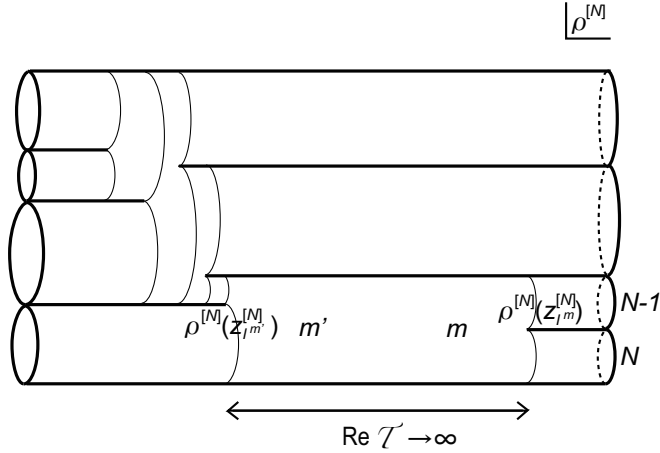}
\par\end{centering}

\caption{The super light-cone diagram for the factorization. 
         \label{fig:factorization}}

\end{figure}

Finally we explore the factorization properties of 
$\Gamma_{\mathrm{super}}$.
Let us consider the super light-cone diagram depicted in 
figure~\ref{fig:factorization}
in the limit $\mathop{\mathrm{Re}}\mathcal{T}\to\infty$. 
Here we attach the superscript
$\left[N\right]$ to the $\rho$ coordinate 
on the super light-cone diagram
to indicate that the diagram is with $N$ external lines. 
We introduce a shorthand notation 
\begin{equation}
\Gamma_{\mathrm{super}}^{\left[N\right]}
   \left(1,2,\ldots,N\right) 
\equiv \Gamma_{\mathrm{super}}
         \left[\ln\left(\left(D\xi^{\left[N\right]}\right)^{2}
                  \left(\bar{D}\bar{\xi}^{\left[N\right]}\right)^{2}
         \right)
         \right]\; ,
\end{equation}
where $1,\ldots,N$ label the external lines. 
Since
$\exp \left[-\Gamma_{\mathrm{super}}^{\left[N\right]}
               \left(1,2,\cdots,N\right)\right]$
can be regarded as the partition function 
corresponding to the super light-cone diagram, 
it should satisfy 
\begin{eqnarray}
\lefteqn{
-\Gamma_{\mathrm{super}}^{\left[N\right]}\left(1,2,\cdots,N\right)
} \nonumber\\
 && \sim 
    {}-\Gamma_{\mathrm{super}}^{\left[N-1\right]}
                   \left(1,2,\cdots,N-2,m'\right)
      -\Gamma_{\mathrm{super}}^{\left[3\right]}
                   \left(m,N-1,N\right)
      + \frac{\mathop{\mathrm{Re}}\mathcal{T}}{\alpha_{m}}\; ,
\label{eq:factorization}
\end{eqnarray}
in the limit $\mathop{\mathrm{Re}}\mathcal{T}\rightarrow\infty$.%
\footnote{Here we consider the factorization in the context of CFT.
          In the context of string field theory, 
         we need to take care of the extra contributions
         coming from the integration measure 
         for the moduli parameters. } 
Here $m,m'$ correspond to the intermediate string. 
It is straightforward to show that 
$\Gamma_{\mathrm{super}}$ given in eq.(\ref{eq:Gammasuper2})
satisfies eq.(\ref{eq:factorization}). 
Eq.(\ref{eq:factorization}) no longer holds,
if we add to $\Gamma_{\mathrm{super}}$
the terms which depend on $\xi_{I}$  but are
independent of $\mathcal{T}_{I}$. 
Therefore we conclude that the right hand side of 
eq.(\ref{eq:Gammasuper2}) gives $-\Gamma_{\mathrm{super}}$. 

%%%%%%%%%%%%%%%%%%%%%%%%%%%%%%%%%%%%%%%%%%%%%%%%%%%%%%%%%%%
\section{Correlation functions\label{sec:Correlation}}

\subsection{One-point function 
        $\left\langle DX^{-} (\mathbf{z}) \right\rangle_{\rho}$}

Using $\Gamma_{\mathrm{super}}$ obtained 
in eq.(\ref{eq:Gammasuper2}),
we can calculate the correlation functions 
and show that the energy momentum tensor (\ref{eq:TXpm}) 
satisfies the desired properties.
%As in the bosonic case~\cite{Baba:2009ns}, 
%we introduce 
%\begin{equation}
%\rho'\left(\mathbf{z}\right) 
%\equiv 
%\sum_{r=0}^{N+1} \alpha_{r}
%  \ln\left(\mathbf{z}-\mathbf{Z}_{r}\right)\; ,
%\quad
%\mbox{with \ $\alpha_{N+1}  =  -\alpha_{0}$}\;,
%\end{equation}
%and accordingly $\xi^{\prime}\left(\mathbf{z}\right)$.
%Then the one-point function 
%$\left\langle DX^{-}\left(\mathbf{z}\right)\right\rangle _{\rho}$
%is obtained as 
%\begin{equation}
%\left\langle DX^{-}\left(\mathbf{Z}_{0}\right)\right\rangle _{\rho}
% =  \left. 2i D_{\mathbf{Z}_{0}}\partial_{\alpha_{0}}\left(-\Gamma_{\mathrm{super}}^{\prime}\right)\right)\right|_{\alpha_{0}=0}\ ,\end{eqnarray*}
%where $\Gamma_{\mathrm{super}}^{\prime}$ denotes $\Gamma_{\mathrm{super}}^{\prime}\left[\ln\left(\left(D\xi^{\prime}\right)^{2}\left(\bar{D}\bar{\xi}^{\prime}\right)^{2}\right)\right]$the
%It is straightforward to calculate the right hand side and
Plugging eq.(\ref{eq:Gammasuper2}) into eq.(\ref{eq:onepoint}), 
we can evaluate the one-point function,
\begin{eqnarray}
\lefteqn{
  \left\langle DX^{-}\left(\mathbf{z}\right)
  \right\rangle _{\rho}
= 2i\frac{d-10}{8}
}\nonumber \\
&& \times D
   \left[\sum_{I}\left\{ 
      \frac{\theta-\tilde{\theta}_{I}}
           {\left(\mathbf{z}-\tilde{\mathbf{z}}_{I}\right)^{4}}
      \frac{13}{2}
      \frac{D\rho}{\left(\partial^{2}\rho\right)^{2}}
         \left(\tilde{\mathbf{z}}_{I}\right)
%   \right.\right.
%\nonumber \\
% &  & \hphantom{\times D\sum_{I}}
%      \quad
%   {}
+\frac{1}{\left(\mathbf{z}-\tilde{\mathbf{z}}_{I}\right)^{3}}
      \frac{4\partial^{2}D\rho D\rho}
           {\left(\partial^{2}\rho\right)^{3}}
        \left(\tilde{\mathbf{z}}_{I}\right)
\right.\right.
\nonumber \\
 &  & \hphantom{\times D\sum_{I}}
      \quad
  {}+ \frac{\theta-\tilde{\theta}_{I}}
           {\left(\mathbf{z}-\tilde{\mathbf{z}}_{I}\right)^{3}}
       \left(\frac{2D\rho}{\left(\partial^{2}\rho\right)^{2}}
             \partial\ln f_{I}
            - \frac{4\partial^{3}\rho D\rho}
                   {\left(\partial^{2}\rho\right)^{3}}\right)
       \left(\tilde{\mathbf{z}}_{I}\right)
\nonumber \\
 &  & \hphantom{\times D\sum_{I}}\quad
   {}+\frac{1}{\left(\mathbf{z}-\tilde{\mathbf{z}}_{I}\right)^{2}}
      \left(\frac{D\rho}{\left(\partial^{2}\rho\right)^{2}}
             \partial D\ln f_{I}
            +\frac{2\partial^{2}D\rho D\rho}
                  {\left(\partial^{2}\rho\right)^{3}}
              \partial\ln f_{I}
      \right.
\nonumber \\
 &  & \hphantom{\times D\sum_{I} \quad
                {}+\frac{1}{\left(\mathbf{z}
                         -\tilde{\mathbf{z}}_{I}\right)^{2}}
               }
   \left.{}+\frac{3}{4\partial^{2}\rho}
      +\frac{13}{6}\frac{\partial^{3}D\rho D\rho}
                        {\left(\partial^{2}\rho\right)^{3}}
      -6\frac{\partial^{3}\rho\partial^{2}D\rho D\rho}
             {\left(\partial^{2}\rho\right)^{4}}
   \right)  \left(\tilde{\mathbf{z}}_{I}\right)
\nonumber \\
 &  & \hphantom{\times D\sum_{I}}\quad
   {}+\frac{\theta-\tilde{\theta}_{I}}
           {\left(\mathbf{z}-\tilde{\mathbf{z}}_{I}\right)^{2}}
      \left(\frac{D\rho}{\left(\partial^{2}\rho\right)^{2}}
               \partial^{2}\ln f_{I}
            -\frac{\partial^{3}\rho D\rho}
                  {\left(\partial^{2}\rho\right)^{3}}
               \partial\ln f_{I}
            -\frac{1}{\partial^{2}\rho}D\ln f_{I}
     \right.
\nonumber \\
 &  & \left.
      \hphantom{\times D\sum_{I} 
                {}+\frac{\theta-\tilde{\theta}_{I}}
                        {\left(\mathbf{z}
                           -\tilde{\mathbf{z}}_{I}\right)^{2}}
               }
      {} +\frac{3}{4} \frac{\partial^{2}D\rho}
                           {\left(\partial^{2}\rho\right)^{2}}
        -\frac{13}{12}\frac{\partial^{4}\rho D\rho}
                           {\left(\partial^{2}\rho\right)^{3}}
        -\frac{\partial^{3}D\rho\partial^{2}D\rho D\rho}
              {6 \left(\partial^{2}\rho\right)^{4}}
        +2\frac{\left(\partial^{3}\rho\right)^{2}D\rho}
               {\left(\partial^{2}\rho\right)^{4}}
       \right)
       \left(\tilde{\mathbf{z}}_{I}\right)
\nonumber \\
 &  & \hphantom{\times D\sum_{I}}\quad
   {}+ \frac{1}{\mathbf{z}-\tilde{\mathbf{z}}_{I}}
     \left(
        \frac{D\rho}{\left(\partial^{2}\rho\right)^{2}}
         \partial^{2}D\ln f_{I}
       + \frac{2\partial^{2}D\rho D\rho}
              {\left(\partial^{2}\rho\right)^{3}}
          \partial^{2}\ln f_{I}
       -\frac{\partial^{3}\rho D\rho}
             {\left(\partial^{2}\rho\right)^{3}}
          \partial D\ln f_{I}
      \right.\nonumber \\
 &  & \hphantom{\times D\sum_{I}\quad
                {}+\frac{1}{\mathbf{z}-\tilde{\mathbf{z}}_{I}}
               }
      {}+\left(\frac{1}{\partial^{2}\rho}
               +\frac{\partial^{3}D\rho D\rho}
                     {\left(\partial^{2}\rho\right)^{3}}
               -\frac{3\partial^{3}\rho\partial^{2}D\rho D\rho}
                     {\left(\partial^{2}\rho\right)^{4}}
         \right) \partial\ln f_{I}
       -\frac{\partial^{2}D\rho}
             {\left(\partial^{2}\rho\right)^{2}}
         D\ln f_{I}
\nonumber \\
 &  & \hphantom{\times D\sum_{I}\quad
                {}+\frac{1}{\mathbf{z}-\tilde{\mathbf{z}}_{I}}
               }
  {}- \frac{3}{4}\frac{\partial^{3}\rho}
                      {\left(\partial^{2}\rho\right)^{2}}
   + \frac{13}{12}\frac{\partial^{4}D\rho D\rho}
                       {\left(\partial^{2}\rho\right)^{3}}
   - \frac{37}{12}\frac{\partial^{4}\rho\partial^{2}D\rho D\rho}
                       {\left(\partial^{2}\rho\right)^{4}}
\nonumber \\
 &  & \hphantom{\times D\sum_{I}\quad
                {}+\frac{1}{\mathbf{z}-\tilde{\mathbf{z}}_{I}}
               }
  \left.
   {}-\frac{25}{6}\frac{\partial^{3}D\rho\partial^{3}\rho D\rho}
                       {\left(\partial^{2}\rho\right)^{4}}
     +8\frac{\left(\partial^{3}\rho\right)^{2}
              \partial^{2}D\rho D\rho}
            {\left(\partial^{2}\rho\right)^{5}}
   \right)  \left(\tilde{\mathbf{z}}_{I}\right)
\nonumber \\
 &  & \hphantom{\times D\sum_{I}}\quad
  \left.
  {}+\frac{\theta-\tilde{\theta}_{I}}
          {\mathbf{z}-\tilde{\mathbf{z}}_{I}}
    \left( \frac{1}{\partial^{2}\rho}\partial D\ln f_{I}
           -\frac{\partial^{2}D\rho}
                 {\left(\partial^{2}\rho\right)^{2}}
               \partial\ln f_{I}
          -\frac{13}{12}\frac{\partial^{3}D\rho}
                             {\left(\partial^{2}\rho\right)^{2}}
          +2\frac{\partial^{3}\rho\partial^{2}D\rho}
                 {\left(\partial^{2}\rho\right)^{3}}
    \right)  \left(\tilde{\mathbf{z}}_{I}\right)
  \right\} \nonumber \\
 &  & \hphantom{\times D}
   \left.
   \quad {}-\frac{1}{2}
      \sum_{r} \frac{1}{\alpha_{r}}
      \left\{ 
         \ln 
           \frac{\mathbf{z}-\tilde{\mathbf{z}}_{I}^{\left(r\right)}}
                  {\mathbf{z}-\mathbf{Z}_{r}}
        -\frac{\theta-\tilde{\theta}_{I}^{\left(r\right)}}
               {(\mathbf{z}
                       -\tilde{\mathbf{z}}_{I}^{(r)}
                )^{2}}
          \frac{D\rho}{\partial^{2}\rho}
             (\tilde{\mathbf{z}}_{I}^{(r)})
        -\frac{1}{\mathbf{z}-\tilde{\mathbf{z}}_{I}^{\left(r\right)}}
         \frac{\partial^{2}D\rho D\rho}
              {\left(\partial^{2}\rho\right)^{2}}
           (\tilde{\mathbf{z}}_{I}^{(r)})
       \right\} \right].
\nonumber
\\
\label{eq:DX-}
\end{eqnarray}

%%%%%%%%%%%%%%%%%%%%%%%%%%%%%%%%%%%%%%%%%
\subsection{Energy momentum tensor}

Behavior of the energy momentum tensor can be deduced from 
eq.(\ref{eq:DX-}).
Using eq.(\ref{eq:corr}), one can find that
\begin{eqnarray}
&&
\left\langle T_{X^\pm}\left(\mathbf{z}\right)
         F\left[X^{+}\right] \right\rangle _{\rho}
  \nonumber \\
 & & \qquad = 
\left\langle 
\left(
\frac{1}{2}
\left(
:DX^+\partial X^-:(\mathbf{z})
+:\partial X^+DX^-:(\mathbf{z})
\right)
-
\frac{d-10}{4}
S\left(\mathbf{z},\boldsymbol{\rho}\right)
\right)
         F\left[X^{+}\right] \right\rangle _{\rho}
  \nonumber \\
 & & \qquad = 
    \left\langle T_{X^\pm}\left(\mathbf{z}\right)
    \right\rangle_\rho
       F\left[-\frac{i}{2}\left(\rho+\bar{\rho}\right)\right]
  \nonumber \\
 & & \qquad \quad \hphantom{=}
   {} -\frac{i}{4}\partial \rho (\mathbf{z})
      \int d^{2}\mathbf{z}^{\prime\prime}
       \frac{\theta^{\prime}-\theta^{\prime\prime}}
            {\mathbf{z}^{\prime}-\mathbf{z}^{\prime\prime}}
       \left.
        \frac{\delta F\left[X^{+}\right]}
             {\delta X^{+}\left(\mathbf{z}^{\prime\prime}\right)}
       \right|_{X^{+}=-\frac{i}{2}\left(\rho+\bar{\rho}\right)}
   \nonumber \\
 & & \qquad \quad \hphantom{=}
   {} -\frac{i}{4}D \rho (\mathbf{z})
      \int d^{2}\mathbf{z}^{\prime\prime}
       \frac{1}
            {\mathbf{z}^{\prime}-\mathbf{z}^{\prime\prime}}
       \left.
        \frac{\delta F\left[X^{+}\right]}
             {\delta X^{+}\left(\mathbf{z}^{\prime\prime}\right)}
       \right|_{X^{+}=-\frac{i}{2}\left(\rho+\bar{\rho}\right)}
      \ , 
\end{eqnarray}
where
\begin{equation}
\left\langle T_{X^{\pm}} \left(\mathbf{z}\right)
\right\rangle _{\rho} 
=  \frac{1}{2}
   \left( -\frac{i}{2}D\rho\left(\mathbf{z}\right)
           \left\langle \partial X^{-}\left(\mathbf{z}\right)
           \right\rangle _{\rho}
         -\frac{i}{2}\partial\rho \left(\mathbf{z}\right)
             \left\langle DX^{-}\left(\mathbf{z}\right)
             \right\rangle _{\rho}
   \right)
  - \frac{d-10}{4}
     S\left(\mathbf{z},\boldsymbol{\rho}\right)\; .
\end{equation}
We can show that
\begin{equation}
\left\langle T_{X^{\pm}}\left(\mathbf{z}\right)
\right\rangle _{\rho} 
  \sim 
  \left\{ 
    \begin{array}{ll}
       \mbox{regular}
         & \quad \left(\mathbf{z}\sim\mathbf{z}_{I}\right)\\
       \mbox{regular}
         & \quad \left(\mathbf{z}\sim\infty\right)\\
         \displaystyle
     \frac{\frac{1}{2}}{\mathbf{z}-\mathbf{Z}_{r}}
            D_{r} \left( - \frac{d-10}{8} \Gamma_{\mathrm{super}}
                  \right)
         + \frac{\theta -\Theta_{r}}{\mathbf{z}-\mathbf{Z}_{r}}
            \partial_{r} \left( -\frac{d-10}{8} 
                                \Gamma_{\mathrm{super}} \right)
         & \quad \left(\mathbf{z}\sim\mathbf{Z}_{r}\right)
     \end{array}
   \right.,
\label{eq:Tbehavior}
\end{equation}
where $D_{r}=\partial_{\Theta_{r}}+\Theta_{r} \partial_{Z_{r}}$
and $\partial_{r}=\partial_{Z_{r}}$.
Thus we see that the energy momentum tensor is regular 
at $\mathbf{z}=\mathbf{z}_{I},\infty$, 
if there are no operator insertions at these points.
Taking into account the definition (\ref{eq:corr-rho}) 
of the correlation function, from 
eq.(\ref{eq:Tbehavior}) for $\mathbf{z} \sim \mathbf{Z}_{r}$
we can read off the OPE
\begin{equation}
T_{X^{\pm}} (\mathbf{z})
e^{-ip^{+}_{r} X^{-}}
  \left( \mathbf{Z}_{r}, \bar{\mathbf{Z}}_{r} \right)
\sim
\frac{1}{\mathbf{z}-\mathbf{Z}_{r}}
\frac{1}{2}
D e^{-ip^{+}_{r} X^{-}}
  \left( \mathbf{Z}_{r}, \bar{\mathbf{Z}}_{r} \right)
+ \frac{\theta - \Theta_{r}}{\mathbf{z}-\mathbf{Z}_{r}}
  \partial e^{-ip^{+}_{r} X^{-}}
  \left( \mathbf{Z}_{r}, \bar{\mathbf{Z}}_{r} \right)~.
\end{equation}
The non-local operator $e^{-ip^{+}_{r} X^{-}}$ therefore 
behaves as a primary field
of weight $0$,
as in the bosonic case \cite{Baba:2009ns}.

%%%%%%%%%%%%%%%%%%%%%%%%%%%%%%%%%%%%%%%%%%%%%%%
\subsection{Super Virasoro algebra}

Substituting eq.(\ref{eq:DX-})
into the first relation in eq.(\ref{eq:corr}),
one can obtain the two-point function
$\left\langle DX^{-} (\mathbf{z}) DX^{-} (\mathbf{z}')
 \right\rangle_{\rho}$.
{}From its singular part for $\mathbf{z} \sim \mathbf{z}'$,
we can deduce the OPE of $DX^{-}$:
\begin{eqnarray}
\lefteqn{
  DX^{-}\left(\mathbf{z}\right)DX^{-}\left(\mathbf{z}'\right)
} \nonumber \\
 &\sim& {}-\frac{d-10}{4}DD^{\prime}
    \left[ \frac{\theta-\theta'}
                {\left(\mathbf{z}-\mathbf{z}'\right)^{3}}
           \frac{3DX^{+}}{\left(\partial X^{+}\right)^{3}}
             \left(\mathbf{z}'\right)
   \right.
\nonumber \\
 &  & \hphantom{{}-\frac{d-10}{4}DD^{\prime}}
    {}+\frac{1}{\left(\mathbf{z}-\mathbf{z}'\right)^{2}}
       \left(\frac{1}{2\left(\partial X^{+}\right)^{2}}
             +\frac{4\partial DX^{+}DX^{+}}
                   {\left(\partial X^{+}\right)^{4}}
       \right) \left(\mathbf{z}'\right)
\nonumber \\
 &  & \hphantom{{}-\frac{d-10}{4}DD^{\prime}}
     {}+\frac{\theta-\theta'}
             {\left(\mathbf{z}-\mathbf{z}'\right)^{2}}
       \left(-\frac{\partial DX^{+}}
                   {\left(\partial X^{+}\right)^{3}}
             -\frac{5\partial^{2}X^{+}DX^{+}}
                   {2\left(\partial X^{+}\right)^{4}}
        \right) \left(\mathbf{z}'\right)
\nonumber \\
 &  & \hphantom{{}-\frac{d-10}{4}DD^{\prime}}
    {}+\frac{1}{\mathbf{z}-\mathbf{z}'}
      \left(-\frac{\partial^{2}X^{+}}
                  {2\left(\partial X^{+}\right)^{3}}
            +\frac{2\partial^{2}DX^{+}DX^{+}}
                  {\left(\partial X^{+}\right)^{4}}
       -\frac{8\partial^{2}X^{+}\partial DX^{+}DX^{+}}
             {\left(\partial X^{+}\right)^{5}}
       \right)   \left(\mathbf{z}'\right)
\nonumber \\
 &  & \hphantom{{}-\frac{d-10}{4}DD^{\prime}}
     {}+\frac{\theta-\theta'}{\mathbf{z}-\mathbf{z}'}
        \left(-\frac{\partial^{2}DX^{+}}
                    {2\left(\partial X^{+}\right)^{3}}
              +\frac{3\partial^{2}X^{+}\partial DX^{+}}
                    {2\left(\partial X^{+}\right)^{4}}
              -\frac{\partial^{3}X^{+}DX^{+}}
                    {2\left(\partial X^{+}\right)^{4}}
         \right.
\nonumber \\
 &  & \hphantom{{}-\frac{d-10}{4}DD^{\prime}
                 \quad+\frac{\theta-\theta'}{\mathbf{z}-\mathbf{z}'}
                }
     \left.\left.
       {}+\frac{\left(\partial^{2}X^{+}\right)^{2}DX^{+}}
               {\left(\partial X^{+}\right)^{5}}
         -\frac{\partial^{2}DX^{+}\partial DX^{+}DX^{+}}
               {\left(\partial X^{+}\right)^{5}}
      \right)\left(\mathbf{z}'\right)
      \right].
\end{eqnarray}
Using this OPE, one can show that 
the energy momentum tensor $T_{X^{\pm}}\left(\mathbf{z}\right)$
satisfies the super Virasoro algebra 
with the central charge $\hat{c}=12-d$:
\begin{eqnarray}
\lefteqn{
T_{X^{\pm}}
  \left(\mathbf{z}\right)T_{X^{\pm}}\left(\mathbf{z}^{\prime}\right)
}\nonumber \\
&& \sim
   \frac{12-d}{4\left(\mathbf{z}-\mathbf{z}^{\prime}\right)^{3}}
  +\frac{\theta-\theta^{\prime}}
        {\left(\mathbf{z}-\mathbf{z}^{\prime}\right)^{2}}
    \frac{3}{2} T_{X^{\pm}}\left(\mathbf{z}^{\prime}\right)
  +\frac{1}{\mathbf{z}-\mathbf{z}^{\prime}}
    \frac{1}{2}DT_{X^{\pm}} \left(\mathbf{z}^{\prime}\right)
  +\frac{\theta-\theta^{\prime}}{\mathbf{z}-\mathbf{z}^{\prime}}
    \partial T_{X^{\pm}}\left(\mathbf{z}^{\prime}\right).
\end{eqnarray}
It follows that
combined with the transverse variables
$X^{i}\left(\mathbf{z},\bar{\mathbf{z}}\right)$,
the total central charge of the system  becomes $\hat{c}=10$. 
This implies that
with the ghost superfields $B\left(\mathbf{z}\right)$
and $C\left(\mathbf{z}\right)$ defined as
\begin{equation}
B(\mathbf{z}) = \beta (z) + \theta b(z)\;,
\qquad
C(\mathbf{z}) = c(z) + \theta \gamma (z) \;,
\end{equation}
it is possible to construct a nilpotent BRST charge
\begin{equation}
Q_{\mathrm{B}}  
=  \oint\frac{d\mathbf{z}}{2\pi i}
   \left[ - C\left(T_{X^{\pm}}-\frac{1}{2}DX^{i}\partial X^{i}
             \right)
          + \left(C\partial C-\frac{1}{4}\left(DC\right)^{2}
            \right)B
   \right].
\end{equation}

\section{Conclusions\label{sec:Conclusions}}

In this paper, we have constructed the supersymmetric generalization
of the theory proposed in Ref.~\cite{Baba:2009ns}. 
Although it is much more complicated 
because of the presence of the odd supermoduli, 
it is possible to generalize the results in Ref.~\cite{Baba:2009ns}. 
We have evaluated $\Gamma_{\mathrm{super}}$, 
which can be used to define and calculate the correlation functions. 
We have shown that the energy momentum tensor is regular 
at $\mathbf{z}=\mathbf{z}_{I}$ and $\infty$
if no operators are inserted there. 
With the transverse variables and ghosts, 
we have shown that it is possible to construct a nilpotent
BRST charge. 

With the BRST charge, now it is possible to construct BRST invariant
worldsheet theory corresponding to 
string theory in noncritical dimensions.
It can be used to dimensionally regularize 
the light-cone gauge superstring field theory \cite{Baba:2009kr}. 
In order to do so, a careful study on the Ramond sector
is also necessary.
We will deal with such an application elsewhere. 

%%%%%%%%%%%%%%%%%%%%%%%%%%%%%%%%%%%%%%%%%%%%%%%%%%%
\section*{Acknowledgements}

N.I. would like to thank the organizers of the 
workshop ``Branes, Strings and Black Holes" at YITP Kyoto, 
for the hospitality, 
where part of this work was done. 
This work was supported in part by 
Grant-in-Aid for Scientific Research~(C) (20540247)
and
Grant-in-Aid for Young Scientists~(B) (19740164) from
the Ministry of Education, Culture, Sports, Science and
Technology (MEXT).

%%%%%%%%%%%%%%%%%%%%%%%%%%%%%%%%%%%%%%%%%%%%%%
%%%%%%%%%%%%%%%%%%%%%%%%%%%%%%%%%%%%%%%%%%%%%%
\appendix

\section{Properties of $\tilde{\mathbf{z}}_{I}$,
         $\mathbf{z}_{I}$ and $\xi_{I}$}
 \label{sec:int-points}

In order to fix the notation,
in this appendix we present  some properties
concerning the interaction points
$\tilde{\mathbf{z}}_{I}, \mathbf{z}_{I}$
and the odd moduli  parameters $\xi_{I}$ introduced in
eq.(\ref{eq:hatrhodef}).
These properties are given in 
Refs.~\cite{Berkovits:1987gp,Aoki:1990yn}.

$w(\mathbf{z})$ introduced in eq.(\ref{eq:wdef})
is useful to derive them.
By the definitions (\ref{eq:hatrhodef}) and (\ref{eq:wdef}),
we have
\begin{equation}
\hat{\rho} (\mathbf{z})
 = \frac{1}{2} \left(w (\mathbf{z})\right)^{2}\;.
\end{equation}
Eq.(\ref{eq:hatrhoexp}) therefore means that
\begin{equation}
w(\mathbf{z}_{I})=0\;,
\qquad
Dw (\mathbf{z}_{I})=0\;.
\end{equation}
Using these equations and the relation
$\eta (\mathbf{z})
 =\frac{Dw}{(\partial w)^{\frac{1}{2}}}(\mathbf{z})$,
one can obtain from eq.(\ref{eq:wdef})
\begin{eqnarray}
&& 
D\rho (\mathbf{z}_{I})
  = \left(\partial w (\mathbf{z}_{I}) \right)^{\frac{1}{2}}
    \xi_{I}\;,
\qquad
\partial \rho (\mathbf{z}_{I})
  = \frac{\partial Dw}{(\partial w)^{\frac{1}{2}}}(\mathbf{z}_{I})
    \xi_{I}\;,
\nonumber\\
&&
\partial D\rho (\mathbf{z}_{I})
 = \frac{1}{2} \frac{\partial^{2}w}{(\partial w)^{\frac{1}{2}}}
                   (\mathbf{z}_{I}) \xi_{I}\;,
\qquad
\partial^{2} \rho (\mathbf{z}_{I})
 =\left( \partial w (\mathbf{z}_{I}) \right)^{2}
  + \left( \frac{\partial^{2}Dw}{(\partial w)^{\frac{1}{2}}}
           - \frac{\partial^{2} w \partial Dw}
                  {(\partial w)^{\frac{3}{2}}}
   \right) (\mathbf{z}_{I})
   \, \xi_{I} \;,
\nonumber\\
&& \partial^{2} D \rho (\mathbf{z}_{I})
   = 2 \partial w \partial Dw (\mathbf{z}_{I})
     + \mathcal{O} (\xi_{I})\;,
\qquad
  \partial^{3} \rho (\mathbf{z}_{I})
     = 3 \partial^{2}w \partial w (\mathbf{z}_{I})
       + \mathcal{O} (\xi_{I})\;,
\nonumber\\
&& \partial^{3} D \rho (\mathbf{z}_{I})
 = 3 \partial^{2} Dw \partial w (\mathbf{z}_{I})
   + 3 \partial^{2}w \partial Dw (\mathbf{z}_{I})
   + \mathcal{O} (\xi_{I})\;.
\end{eqnarray}
These relations yield
\begin{equation}
\xi_{I} = \frac{D\rho}{(\partial^{2} \rho)^{\frac{1}{4}}}
           (\mathbf{z}_{I})\;,
\label{eq:xiIdef}
\end{equation}
and
\begin{equation}
\partial \rho (\mathbf{z}_{I})
 -\frac{1}{2} 
  \frac{\partial^{2}D\rho D\rho}{\partial^{2}\rho}
    (\mathbf{z}_{I})=0\;,
\qquad
\partial D \rho (\mathbf{z}_{I})
 - \frac{1}{6} \frac{\partial^{3}\rho D\rho}
                    {\partial^{2} \rho} (\mathbf{z}_{I})
 =0~.
\label{eq:zI-another}
\end{equation}

Taylor expanding eq.(\ref{eq:int-pt-tilde})
around $\tilde{\mathbf{z}}_{I} \sim \mathbf{z}_{I}$,
one obtains equations for
$\tilde{\mathbf{z}}_{I}-\mathbf{z}_{I}$
and $\tilde{\theta}_{I}-\theta_{I}$. 
Solving them, we have
\begin{equation}
\tilde{\mathbf{z}}_{I} - \mathbf{z}_{I}
    = -\frac{\partial \rho}{\partial^{2} \rho}
       (\mathbf{z}_{I})
   = -\frac{\partial Dw}{(\partial w)^{\frac{5}{2}}}
      (\mathbf{z}_{I})
    \, \xi_{I}\;,
\quad
\tilde{\theta}_{I} - \theta_{I}
 = -\frac{\partial D\rho}{\partial^{2}\rho} (\mathbf{z}_{I})
 = - \frac{1}{2}
     \frac{\partial^{2}w}{(\partial w)^{\frac{5}{2}}}
             (\mathbf{z}_{I}) \, \xi_{I}~.
\label{eq:difference}
\end{equation}
We note that $\tilde{\mathbf{z}}_{I}-\mathbf{z}_{I}$
and $\tilde{\theta}_{I} - \theta_{I}$
are proportional to $\xi_{I}$.
This yields
\begin{equation}
D\rho (\mathbf{z}_{I})=D\rho (\tilde{\mathbf{z}}_{I})\;,
\qquad
\xi_{I} = \frac{D\rho}{(\partial^{2} \rho)^{\frac{1}{4}}}
           (\mathbf{z}_{I})
       = \frac{D\rho}{(\partial^{2} \rho)^{\frac{1}{4}}}
           (\tilde{\mathbf{z}}_{I})\;,
\label{eq:xiIdef2}
\end{equation}
as well as eq.(\ref{eq:rhozIeq}).

%%%%%%%%%%%%%%%%%%%%%%%%%%%%%%%%%%%%%%%%%
\section{Another Expression for $\Gamma_{\mathrm{super}}$}
\label{sec:otherGamma}

In this appendix, we recast $\Gamma_{\mathrm{super}}$
given in eq.(\ref{eq:Gammasuper2}) into an convenient expression 
to show that this is consistent with the results in 
Refs.~\cite{Berkovits:1985ji,Berkovits:1987gp}.

{}From eqs.(\ref{eq:omegafact}) and (\ref{eq:fI-1}),
one can find that
\begin{eqnarray}
\ln \omega (\mathbf{z})
 &=& \ln A - \sum_{r} \ln (\mathbf{z}-\mathbf{Z}_{r})
     + \sum_{I} P_{I} \ln (\mathbf{z}-\tilde{\mathbf{z}}_{I})~,
\nonumber\\
\ln f_{I} (\mathbf{z})
 &=& \ln A - \sum_{r} \ln (\mathbf{z}-\mathbf{Z}_{r})
     + \sum_{J\neq I} P_{J} 
       \ln (\mathbf{z} -\tilde{\mathbf{z}}_{J})~,
\label{eq:lnfI}
\end{eqnarray}
where $P_{I}$ is defined in eq.(\ref{eq:W-PI}).
Using eq.(\ref{eq:lnfI}) and  the relation
$\lim_{\mathbf{z} \rightarrow \mathbf{Z}_{r}}
   (\mathbf{z}-\mathbf{Z}_{r}) \omega (\mathbf{z})
  = \alpha_{r}$,
we can rewrite $W_{\mathrm{super}}$
given in eq.(\ref{eq:W-PI}) into
\begin{equation}
-W_{\mathrm{super}}
 =  - \frac{1}{2} \sum_{I} F_{I} (\tilde{\mathbf{z}}_{I})
      -\frac{1}{2} \sum_{r} \ln \alpha_{r}
      +\ln A~,
\label{eq:Wappend1}
\end{equation}
where $F_{I} (\mathbf{z})$ is defined in eq.(\ref{eq:defFI}).
Using eqs.(\ref{eq:omegadef}) and (\ref{eq:fI-1}),
one can get
\begin{equation}
\ln f_{I} (\mathbf{z})
 = \ln \frac{\partial \rho (\mathbf{z})}
            {\mathbf{z}-\tilde{\mathbf{z}}_{I}}
  - \frac{\partial D\rho D\rho}{(\partial \rho)^{2}}(\mathbf{z})
  + \frac{1}{\mathbf{z}-\tilde{\mathbf{z}}_{I}}
    \frac{\partial^{2}D\rho D\rho}
         {(\partial^{2} \rho)^{2}} (\tilde{\mathbf{z}}_{I})
  + \frac{\theta -\tilde{\theta}_{I}}
         {(\mathbf{z}-\tilde{\mathbf{z}}_{I})^{2}}
    \frac{D\rho}{\partial^{2} \rho}
          (\tilde{\mathbf{z}}_{I})~.
\end{equation}
This yields
\begin{eqnarray}
\ln f_{I} (\tilde{\mathbf{z}}_{I})
 &=& 
     \ln \partial^{2} \rho (\tilde{\mathbf{z}}_{I})
     -\frac{1}{2}
       \frac{\partial^{3}D\rho D\rho}{(\partial^{2} \rho)^{2}}
               (\tilde{\mathbf{z}}_{I})
       + \frac{\partial^{3}\rho \partial^{2}D\rho D\rho}
              {(\partial^{2}\rho)^{3}}
                (\tilde{\mathbf{z}}_{I})\;,
\nonumber\\
\partial \ln f_{I} (\tilde{\mathbf{z}}_{I})
 &=& \frac{1}{2} \frac{\partial^{3}\rho}
                      {\partial^{2} \rho} 
                         (\tilde{\mathbf{z}}_{I})
     +
     \left(
     \mbox{terms proportional to $D\rho (\tilde{\mathbf{z}}_I)$}
     \right),
%    -\frac{1}{6}
%     \frac{\partial^{4}D\rho D\rho}{(\partial^{2}\rho)^{2}}
%                 (\tilde{\mathbf{z}}_{I})
%    + \frac{1}{2}
%      \frac{\partial^{3}\rho \partial^{3}D\rho D\rho}
%           {(\partial^{2} \rho)^{3}}
%                (\tilde{\mathbf{z}}_{I})
%   + \frac{1}{3} \frac{\partial^{4}\rho \partial^{2}D\rho D\rho}
%                       {(\partial^{2} \rho)^{3}}
%                          (\tilde{\mathbf{z}}_{I})
%  \nonumber\\
% && {}-\frac{3}{4} \frac{(\partial^{3}\rho)^{2} \partial^{2}D\rho
%                       D\rho}
%                      {(\partial^{2}\rho)^{4}}
%                  (\tilde{\mathbf{z}}_{I}) \;,
  \nonumber\\
\partial D \ln f_{I}  (\tilde{\mathbf{z}}_{I})
 &=& \frac{5}{6} 
     \frac{\partial^{3} D\rho}{\partial^{2} \rho}
          (\tilde{\mathbf{z}}_{I})
    - \frac{\partial^{3}\rho \partial^{2}D\rho}
            {(\partial^{2} \rho)^{2}}
              (\tilde{\mathbf{z}}_{I})
              +
              \left(
              \mbox{terms proportional to
                     $D\rho (\tilde{\mathbf{z}}_I)$}
              \right).
%   -\frac{1}{12} \frac{\partial^{5}\rho D\rho}
%                      {\partial^{2} \rho}
%               (\tilde{\mathbf{z}}_{I})
%  + \frac{1}{3} 
%      \frac{\partial^{3}\rho \partial^{4}\rho D\rho}
%           {(\partial^{2} \rho)^{3}}
%              (\tilde{\mathbf{z}}_{I})
%\nonumber\\
%&&{}  -\frac{1}{4} \frac{(\partial^{3}\rho)^{3} D\rho}
%                    {(\partial^{2}\rho)^{4}}
%                       (\tilde{\mathbf{z}}_{I})~.
\label{eq:lnfIappned}
\end{eqnarray}
Substituting these equations into eq.(\ref{eq:Wappend1}),
we obtain
\begin{equation}
-W_{\mathrm{super}} 
=  \frac{1}{2} \sum_{I}
    \ln \left(\partial^{2} \rho 
               -\frac{4}{3} \frac{\partial^{3}D\rho D\rho}
                                 {\partial^{2} \rho}
               + \frac{5}{2} 
                 \frac{\partial^{3}\rho \partial^{2}D\rho D\rho}
                      {(\partial^{2} \rho)^{2}}
        \right)  (\tilde{\mathbf{z}}_{I})
  -\frac{1}{2} \sum_{r} \ln \alpha_{r}
  +\ln A~,
\end{equation}
and thus
\begin{eqnarray}
-\Gamma_{\mathrm{super}}
 &=& -\frac{1}{4} \sum_{I}
     \ln \left( \partial^{2} \rho
             - \frac{5}{3} \frac{\partial^{3}D\rho D\rho}
                                {\partial^{2}\rho}
            +3 \frac{\partial^{3}\rho \partial^{2}D\rho D\rho}
                    {(\partial^{2}\rho)^{3}}
         \right) (\tilde{\mathbf{z}}_{I})
\nonumber\\
  && \quad {}-\frac{1}{2} \sum_{r} \bar{N}^{rr}_{00}
  -\frac{1}{2} \sum_{r} \ln \alpha_{r}
  + \ln A
\nonumber\\
 && {}+\mbox{c.c.}~.
\end{eqnarray}
We set $Z_{N}=\infty$ using the superprojective invariance
of $\Gamma_{\mathrm{super}}$. 
Then $e^{-\Gamma_{\mathrm{super}}}$ 
can be compared with eq.(4.10) in Ref.~\cite{Berkovits:1987gp}. 
Taking into account the fact that 
a factor $e^{\bar{N}^{rr}_{00}}$ comes from $\hat{\epsilon}_r$, 
eq.(4.10) in Ref.~\cite{Berkovits:1987gp} coincides with 
$e^{-\Gamma_{\mathrm{super}}}$.

%%%%%%%%%%%%%%%%%%%%%%%%%%%%%%%%%%%%%%%%%%%%%%%
\bibliographystyle{utphys}
\bibliography{SFT}

\end{document}